\documentclass[graybox, secnum]{svmult}

\usepackage{amssymb,amsmath}

\usepackage{mathtools}

\usepackage{mathptmx}       
\usepackage{helvet}         
\usepackage{courier}        
\usepackage{type1cm}        
\usepackage{makeidx}         
\usepackage{graphicx}        
\usepackage{multicol}        
\usepackage[bottom]{footmisc}
\usepackage{hyperref}        
\usepackage{soul}            
\hypersetup{colorlinks=true,urlcolor=blue}
\makeindex             

\usepackage[numbers,sort,compress]{natbib}

\begin{document}
\title*{11D Supergravity and Hidden Symmetries}
\author{Henning Samtleben}
\institute{Henning Samtleben \at ENSL, CNRS, Laboratoire de physique, F-69342 Lyon, Institut Universitaire de France (IUF), France, \email{henning.samtleben@ens-lyon.fr}}
%
%
\maketitle

\vspace*{-3cm}

\abstract{We review the structure of maximal $D=11$ and $D=10$ supergravities.
Upon dimensional reduction, these theories give rise to the unique maximal supergravities in all lower spacetime dimensions $D<10$.
In $D$ dimensions, maximal supergravity exhibits the exceptional global symmetry group E$_{11-D}$,
part of which is realized as hidden symmetries and only manifest after proper dualization of the fields.
We also briefly review the reformulation of $D=11$ supergravity as an exceptional field theory which renders
the appearance of hidden symmetries manifest.
}

\bigskip
\bigskip

\noindent
Invited chapter for {\em Handbook of Quantum Gravity} (Eds. C. Bambi, L. Modesto and I.L. Shapiro, Springer Singapore, expected in 2023).

\vspace*{1cm}

\section*{Keywords} 
Supersymmetry, Supergravity, Higher dimensions, 
Compactification, Kaluza-Klein theory, 
Gauge symmetry, Hidden symmetries, Exceptional groups, Exceptional field theory


\section{Introduction}

The interest to study supergravity theories in higher dimensions $D>4$, and their dimensional reduction is (at least) twofold. On the one hand supergravity theories appear as the low-energy effective action for string theories, which generically live in higher dimensions. 
On the other hand, from a purely four-dimensional point of view, the dimensional reduction of higher-dimensional supergravities can be considered as a powerful technique in order to construct extended $D=4$ supergravity theories, i.e., supergravities with ${\cal N}>1$ supercharges.

Under dimensional reduction of a minimally supersymmetric theory (i.e., a theory with a single supercharge), 
the spinor supercharge in general breaks into a number of lower-dimensional spinors, 
according to the fundamental spinor representations of the corresponding Poincar\'e groups.\footnote{
To get the correct counting, the reality conditions of spinors in the different dimensions of spacetime
have to be taken into account, see e.g., \cite{Strathdee:1986jr}.} 
Minimal supersymmetry in higher dimensions thus induces extended supersymmetry in the reduced theory.
For globally supersymmetric theories, this led to the direct construction of maximally supersymmetric $D=4$, ${\cal N}=4$ Yang-Mills theory  upon dimensional reduction of the ten-dimensional minimal ${\cal N}=1$ super Yang-Mills theory~\cite{Brink:1976bc}.

For supergravity, eleven is the highest dimension in which the minimal 
supersymmetric extension of the Poincar\'e algebra allows for a supermultiplet of fields of spin smaller or equal to 2~\cite{Nahm:1977tg}.
For fields with spin larger than 2, interacting theories with a finite number of fields in general do not exist.
Eleven is thus the highest dimension in which supergravity can be constructed.
The associated interacting theory is unique and has been found by Cremmer, Julia, and Scherk~\cite{Cremmer:1978km}.
Upon dimensional reduction to $D=4$ dimensions, this theory gives rise to maximally supersymmetric ${\cal N}=8$ supergravity~\cite{Cremmer:1978ds,Cremmer:1979up}. This is arguably the most remarkable extension of $D=4$ Einstein gravity 
due to its high degree of symmetry and the finiteness properties of its 
higher loop amplitudes~\cite{Bern:2007hh,Bern:2009kd,Arkani-Hamed:2008owk,Green:2010sp,Vanhove:2010nf,Bjornsson:2010wm,Pioline:2015yea,Bern:2018jmv}.

The detour via eleven dimensions has in fact been an indispensable tool in the construction of this theory~\cite{Cremmer:1978ds,Cremmer:1979up}, 
in particular in order to determine
the complicated non-linear interactions between its 70 scalar fields, which had proven an extremely challenging task 
within the $D=4$ perturbative approach~\cite{deWit:1977fk}. The other key ingredient in the construction of ${\cal N}=8$ supergravity 
has been the seminal observation
that the theory admits an unexpectedly large global symmetry group, the exceptional group ${\rm E}_{7(7)}$~\cite{Cremmer:1978ds,Cremmer:1979up}.\footnote{The subscript in parentheses in
this notation specifies the particular real form of the group: for ${\rm E}_{7(7)}$, the associated Lie algebra has 70 non-compact generators and 63 compact generators
with the latter spanning the compact Lie algebra $\mathfrak{su}(8)$.}
While part of these global symmetries can be understood from the gauge symmetries of the theory's higher-dimensional ancestor,
a considerable part of the exceptional symmetries has no direct higher-dimensional interpretation and is often referred to as {\em hidden symmetries}.
Their presence is essential for the realization of the exceptional symmetry group which in turn allows to organize the couplings of the theory in a
remarkably compact way. For example, the scalar sector of the theory is most concisely described as a non-linear sigma model on the coset space
${\rm E}_{7(7)}/{\rm SU}(8)$.

\begin{figure}[tb]
   \centering
   \includegraphics[width=6cm]{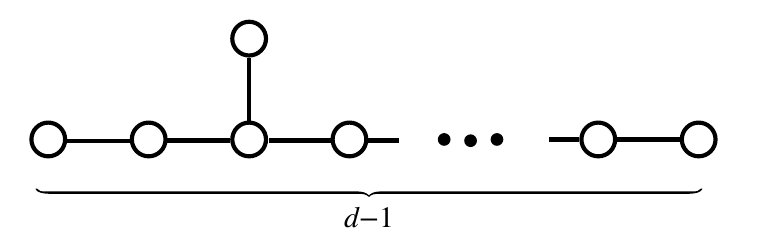}
   \caption{{\small The Dynkin diagram of the exceptional Lie algebra $\mathfrak{e}_d$.}}
   \label{fig:dynkin}
\end{figure}

Subsequently, the presence of hidden symmetries and the 
appearance of exceptional global symmetry groups in maximal supergravity were recognized as 
part of a general pattern that has been dubbed the {\em silver rules of supergravity}~\cite{Julia:1980gr,Julia:1982gx,Julia:1982tm,Julia:1997cy}.
Maximal supergravity in $D=11-d$ dimensions exhibits a global
symmetry group ${\rm E}_{d(d)}$, realizing the series of exceptional Lie groups in the Dynkin classification,
with the Dynkin diagram of the associated algebra given in Figure~\ref{fig:dynkin}.
For small values of $d$, the exceptional series degenerates into the classical Lie groups
\begin{equation}
{\rm E}_{5(5)} \simeq {\rm SO}(5,5)\;,\quad
{\rm E}_{4(4)} \simeq {\rm SL}(5)\;,\quad
{\rm E}_{3(3)} \simeq {\rm SL}(3) \times {\rm SL}(2)
\;,
\end{equation}
as can be found from properly extrapolating the general Dynkin diagram.
Let us also note that discrete versions ${\rm E}_{d(d)}(\mathbb{Z})$ of the exceptional symmetry groups of supergravity
survive in the toroidal compactification of the full string theories~\cite{Hull:1994ys}.

In this chapter, we review the structure of maximal supergravity in eleven dimensions, its dimensional reduction,
and the appearance of hidden symmetries. To this end, we first review in section~\ref{sec:maximal}
in some detail the field content and dynamics of the maximal
supergravities in ten and eleven dimensions. In section~\ref{sec:toroidal}, we discuss the toroidal compactification of these theories
to lower-dimensional maximal supergravities. In particular, we determine the {\em geometric} symmetries of the lower-dimensional
theories, i.e., the global symmetries that descend from particular diffeomorphism and gauge transformations in higher dimensions.
Section~\ref{sec:hidden} then reviews the appearance of {\em hidden} global symmetries in lower dimensions. The central example is
$D=4$, ${\cal N}=8$ supergravity with its global symmetry group enhanced to the full exceptional group ${\rm E}_{7(7)}$.
Finally, in section~\ref{sec:exft}, we review the formulation of 11D supergravity as an {\em exceptional field theory}~\cite{Hohm:2013pua},
which highlights the role of the full exceptional group in the full 11D supergravity before dimensional reduction.

Throughout this chapter, our discussion of hidden symmetries will mostly be 
restricted to the bosonic sectors of the supergravity theories.
Although the fermionic field content and couplings are at the very origin of all these theories, the symmetry enhancement 
and the appearance of the exceptional symmetry groups can be realized and studied entirely within their bosonic sectors.
In particular, even in presence of fermions, the exceptional global symmetry algebras do not extend to larger superalgebras.

\section{Maximal supergravity in $D=11$ and $D=10$ dimensions}
\label{sec:maximal}

\subsection{Field content}

The highest-dimensional supergravity theory lives in eleven spacetime
dimensions and was constructed by Cremmer, Julia, and Scherk~\cite{Cremmer:1978km}.
Its field content is given by the lowest massless representation of the 
supersymmetry algebra
\begin{equation}
\{ Q_\alpha, Q_\beta \} ~=~ 
2 P_M \left( \Gamma^M \Gamma^0 \right)_{\alpha\beta} \,,
\label{susyalg11}
\end{equation} 
where $Q_\alpha$ are the $32$ independent real supercharges in 
eleven-dimensional Minkowski spacetime.
$P_M$ with $M=0, 1, \dots, 10$, is the $D=11$ momentum, and $\Gamma^M$
are the $\mathfrak{so}(1,10)$ gamma matrices.
For massless states which fulfill $P_M P^M=0$, the 
 r.h.s.\ of (\ref{susyalg11}) describes a projector of half-maximal rank
in spinor space. It follows that only 16 out of the 32 supercharges
act non-trivially on massless states and satisfy the Clifford algebra ${\cal C}_{16}$,
which admits a $2^8=256$-dimensional irreducible representation.
The spacetime interpretation of these states is inferred from the embedding of the massless little group
\begin{equation}
{\cal C}_{16}  \supset \mathfrak{so}(16) \supset 
 \mathfrak{so}(9)
\,,
\end{equation} 
according to
\begin{equation}
256 ~\longrightarrow~ 128_s+128_c ~\longrightarrow~ 44+84+128
\;.
\label{169}
\end{equation} 
The $44$ corresponds to the symmetric traceless product of two vectors:
these are the degrees of freedom of a massless spin-2 field, the graviton $G_{MN}$.
The $84=\binom{9}{3}$ on the other hand counts the degrees of freedom 
a totally antisymmetric massless 3-form field $C_{KMN}$,
i.e., a field with local tensor gauge symmetry
\begin{eqnarray}
\delta C_{KMN} &=& 3\,\partial_{[K} \Lambda_{MN]}\;.
\label{varC3}
\end{eqnarray}
The 128 finally corresponds to the degrees of freedom of a massless spin-$3/2$ field
in eleven dimensions, the gravitino $\Psi_M$.
The full 11D supergravity multiplet thus is given by
\begin{equation}
\left\{G_{MN}, \psi_M, C_{KMN}\right\}  . 
\label{11Dmultiplet}
\end{equation}
The interacting theory has been constructed in~\cite{Cremmer:1978km}
and is reviewed in section~\ref{subsec:11D}.
The same reasoning shows why supergravity theories do not exist
beyond eleven dimensions: repeating the above analysis for, say, a twelve-dimensional
spacetime with 64 supercharges yields a minimal field content that includes fields with spin 
larger than two. No consistent interacting theory for such fields can be constructed
(unless infinitely many fields are included).

A similar analysis yields the field content of $D=10$ supergravity.
In this case, the smallest massless representation descends from the Clifford algebra 
${\cal C}_{8}$ and comprises $2^4=16$ states. 
In analogy to (\ref{169}), they transform as vector and spinor under the little group ${\rm SO}(8)$
\begin{eqnarray}
16  ~\longrightarrow~ 8_v + 8_s
\;,
\label{vector10}
\end{eqnarray}
counting degrees of freedom of a ten-dimensional massless vector and a matter fermion. This is the minimal
${\cal N}=1$ vector multiplet in ten dimensions. The higher massless multiplets can be found by tensoring 
(\ref{vector10}) with the fundamental representations of ${\rm SO}(8)$. This results in the ${\cal N}=1$
supergravity multiplet and two inequivalent gravitino multiplets:
\begin{eqnarray}
{\rm sugra} &:& 
8_v \otimes \left(8_v + 8_s\right) ~=~
1+28+35_v + 8_c+ 56_c \;,
\nonumber\\
\mbox{gravitino A} &:& 
8_c \otimes \left(8_v + 8_s\right) ~=~
8_v+56_v+ 8_s+ 56_s\;,
\nonumber\\
\mbox{gravitino B} &:& 
8_s \otimes \left(8_v + 8_s\right) ~=~
1+28+35_s + 8_c+ 56_c \;.
\label{D10AB}
\end{eqnarray}
Similar to (\ref{169}), the physical field content of these multiplets may be inferred from the representations of the little group.
The ${\cal N}=1$ supergravity multiplet carries the metric, a scalar field (the dilaton) and 
an antisymmetric 2-form together with a gravitino and a matter fermion. The interacting theory exists,
and upon coupling to the vector multiplet (\ref{vector10})
in the adjoint representation of the gauge group, this describes the low-energy effective theory of
the heterotic string \cite{Gross:1984dd}.

The first gravitino multiplet in (\ref{D10AB}) carries a spacetime vector and
a  3-form together with a gravitino and a matter fermion of chirality opposite to the fermions of the supergravity multiplet.
Instead, the second gravitino multiplet carries another scalar (the axion) and a 2-form together
with a self-dual 4-form (the $35_s$ in (\ref{D10AB})), 
see~(\ref{FF55}) below. Its fermions are a gravitino and a matter fermion of 
the same chirality as the fermions of the supergravity multiplet. Coupling of a massless
gravitino multiplet requires supersymmetry enhancement. The two resulting theories exist as maximally 
supersymmetric interacting theories and are denoted as the non-chiral ${\cal N}=(1,1)$ type IIA and 
the chiral ${\cal N}=(2,0)$ type IIB theory, respectively. 
They are the low-energy effective theories for the massless spectrum of IIA and IIB strings, respectively.
The type IIA theory can be obtained by compactifying 11D supergravity on a circle $S^1$ as will be discussed
in section~\ref{subsec:IIA}. In contrast, the type IIB
supergravity does not have a higher-dimensional origin and has been constructed in~\cite{Schwarz:1983wa,Schwarz:1983qr,Howe:1983sra}.
We will review it in section~\ref{subsec:IIB}.

\subsection{11D supergravity}
\label{subsec:11D}

The action of eleven-dimensional supergravity with
field content given in (\ref{11Dmultiplet}) has been constructed in~\cite{Cremmer:1978km}.
To quadratic order in the fermions, its Lagrangian extends the standard combination of Einstein-Hilbert and Rarita-Schwinger term
\begin{eqnarray}
{\cal L}_0[{E,\psi}] &=& 
 |E|\, R[\omega]
-\frac12\, |E| \bar\psi_K \Gamma^{KMN} D[\omega]_M \psi_N \;,
\label{111}
\end{eqnarray}
by a kinetic, a topological, and a fermionic 
interaction term for the antisymmetric 3-form field $C_{KMN}$, given by
\begin{eqnarray}
{\cal L}_C[{E,C,\psi}] &=&
-\frac1{48} |E|  F_{KLMN} F^{KLMN}  
+ \frac{1}{144^2} \, \varepsilon^{N_0N_1\dots N_{10}} 
\,F_{N_0\dots N_3} \,F_{N_4\dots N_7} \,C_{N_8 N_{9} N_{10}}   \nonumber\\ 
&&{}
+ \frac{1}{192} |E| \Big(\bar\psi_P \Gamma^{KLMNPQ} 
\psi_Q + 12 \,\bar\psi^K  \Gamma^{LM} \psi^N\Big) F_{KLMN} 
\;.
\label{112}
\end{eqnarray}
Here, $|E|$ denotes the determinant of the eleven-bein $E_M{}^A$, related to the metric as
\begin{eqnarray}
G_{MN}=E_M{}^A E_N{}_A\;,
\end{eqnarray}
with flat Lorentz indices $A$.\footnote{Throughout this chapter, we use spacetime signature $(-++\dots+)$.} 
The gauge invariant abelian field strength is defined as
$F_{KLMN}=4\,\partial_{[K}C_{LMN]}$.
The totally antisymmetric (numerical) $\varepsilon^{N_0N_1\dots N_{10}}$ is the Levi-Civita density.
The Chern-Simons term $F\wedge F\wedge C$ is invariant under tensor gauge transformations
(\ref{varC3}) up to a total derivative. The appearance of such topological terms is 
a generic feature in higher-dimensional supergravity theories.
The full 11D supergravity Lagrangian is given by
\begin{eqnarray}
{\cal L}_{\rm 11D} &=& {\cal L}_0 + {\cal L}_C + {\cal L}_{\psi^4}
\;,
\label{LD11}
\end{eqnarray}
where the quartic fermion terms of ${\cal L}_{\psi^4}$ can be formally absorbed into 
an appropriate modification of the spin connection $\omega$ and the 4-form field strength im 
the $\psi^2$ terms in (\ref{111}), (\ref{112}).
The full Lagrangian (\ref{LD11}) is invariant under the supersymmetry transformations
\begin{eqnarray}
\delta E_M{}^A &=& \frac14\,\bar\epsilon\Gamma^A \psi_M \;,\qquad
\delta C_{KLM} =  \frac34\,\bar\epsilon\Gamma_{[KL} \psi_{M]} \;,
\nonumber\\
\delta \psi_M &=& D[\omega]_M \epsilon-\frac1{288}\,F^{KLPQ} \,\Gamma_{MKLPQ} \, \epsilon
+\frac1{36} \,F_{MNPQ} \,\Gamma^{NPQ}  \epsilon
\;,
\end{eqnarray}
given up to cubic terms in the fermions. In turn, supersymmetry uniquely fixes all the terms in (\ref{LD11})
and in particular requires the presence of the topological Chern-Simons term in (\ref{112}).

Let us still point out two interesting properties of the Lagrangian (\ref{LD11}), which will be important
for the appearance of hidden symmetries. First, the field equations descending from the Lagrangian (\ref{LD11}),
scale homogeneously under the following transformation
\begin{eqnarray}
\delta E_M{}^A &=& \zeta\,E_M{}^A\;,\quad
\delta C_{KMN}~=~ 3\zeta\,C_{KMN}\;,\quad
\delta \psi_M~=~ \frac\zeta2\,\psi_M
\;,
\label{trombone}
\end{eqnarray}
with constant $\zeta$.
In general spacetime dimension $D$, this so-called {\em trombone symmetry}
is not a symmetry  of the action but rescales the Lagrangian as
\begin{eqnarray}
\delta {\cal L} &=& (D-2)\,\zeta\,{\cal L}
\;.
\end{eqnarray}
It still plays an important role, e.g.,
among the spectrum-generating symmetries for the fundamental BPS
solutions~\cite{Cremmer:1997xj}. 
The trombone symmetry is present in all two-derivative supergravity theories, but is in general 
broken by higher-order corrections.

Second, the field equations for the 3-form $C_{KLM}$
may be used as integrability relations which ensure the consistency of the definition of an antisymmetric 6-form potential
$C_{N_1\dots N_6}$ by means of the first-order duality equation
\begin{equation}
F_{N_1N_2\dots N_7}  
=
-\frac1{24}\,|E|\,\varepsilon_{N_1N_2 \dots N_7M_1M_2M_3M_4} F^{M_1M_2M_3M_4}  +~\mbox{fermions}
\,.
\label{F47}
\end{equation}
Here, the 7-form field strength is defined as
\begin{equation}
F_{N_1N_2\dots N_7} =
7\,\partial_{[N_1} C_{N_2N_3\dots N_7]} + \frac{35}{2}\,C_{[N_1N_2N_3} \,F_{N_4N_5N_6N_7]}
\;,
\label{defF7}
\end{equation}
with a non-trivial Bianchi identity
\begin{equation}
8\,\partial_{[N_1} F_{N_2N_3\dots N_8]} =
 35\,F_{[N_1N_2N_3N_4} \,F_{N_5N_6N_7N_8]}
\;.
\label{Bianchi7}
\end{equation}
Indeed, hitting (\ref{F47}) with another external derivative and using (\ref{Bianchi7}), 
this equation reduces to the field equations
for the 3-form $C_{KMN}$, obtained by variation of (\ref{112}). In particular, the contribution from the Chern-Simons term in (\ref{112}) requires the
second term in (\ref{defF7}). As a consequence, the 6-form $C_{N_1\dots N_6}$ transforms 
non-trivially under the gauge transformations (\ref{varC3}). Specifically, its field strength (\ref{defF7}) 
is invariant under the gauge transformations
\begin{equation}
\delta C_{N_1N_2\dots N_6} =
6\,\partial_{[N_1} \Lambda_{N_2N_3\dots N_6]}
 - 30\,  C_{[N_1N_2N_3}\,\partial_{N_4} \Lambda_{N_5N_6]} 
 \;,
 \label{varC6}
\end{equation}
with a new gauge parameter $\Lambda_{N_1\dots N_5}$, and $\Lambda_{MN}$ from (\ref{varC3}).
This shows that the full algebra of 11D gauge symmetries is actually non-abelian 
\begin{equation}
[\delta_{\Lambda_1} , \delta_{\Lambda_2}] = \delta_{\Lambda_{12}}\;,
\label{gaugealgebra11D}
\end{equation}
with the commutator of two 3-form gauge transformations (\ref{varC3}), (\ref{varC6}), resulting in a 6-form gauge transformation with parameter
\begin{equation}
 \Lambda_{12,N_1N_2\dots N_5} = 15 \,\Lambda_{2,[N_1N_2}\,\partial_{\vphantom{[}N_3} \Lambda_{1,N_4N_5]}
-15 \,\Lambda_{1,[N_1N_2}\,\partial_{\vphantom{[}N_3} \Lambda_{2,N_4N_5]}
\;.
\label{gaugealgebra11D2}
\end{equation}

The introduction of dual fields, defined by first-order duality equations such as (\ref{F47}) is a general feature of supergravity theories. 
In general spacetime dimension~$D$, the field equations 
for a given $p$-form allow the introduction of its dual form of degree $(D-p-2)$, in terms of which Bianchi identities and equations of motion
become exchanged. As a consequence, supergravity theories may admit different Lagrangian formulations, which are on-shell equivalent only 
after relating their fields by means of duality equations such as (\ref{F47}).\footnote{Specifically, in dimensional reductions from 11D supergravity,
the lower-dimensional duality equations are precisely obtained by dimensional reduction of (\ref{F47}).}
In particular, there is always a Lagrangian formulation of the theory in which all forms are dualized to the lowest possible degree.
It is in this form that the largest global symmetry group becomes visible, as we shall see below.

\subsection{IIA supergravity}
\label{subsec:IIA}

Dimensional reduction of 11D supergravity on a circle $S^1$ yields maximal type IIA supergravity
in ten dimensions~\cite{Giani:1984wc,Campbell:1984zc}. Explicitly, the reduction amounts to imposing independence of all fields
on the eleventh coordinate
\begin{equation}
\partial_{10} G_{MN} = 0 = \partial_{10} C_{KMN}\;,
\label{d11}
\end{equation}
together with a standard Kaluza-Klein ansatz for the eleven-dimensional metric\footnote{
Compared to the general Kaluza-Klein parametrization (\ref{vielbeinreduced}) given below, this ansatz uses a rescaled 
dilaton $\phi\rightarrow\frac23\,\phi$ in order to match some standard conventions of IIA supergravity.
}
\begin{equation}
ds^2_{(11)} = {\rm e}^{-\phi/6}\,g_{\mu\nu}\,dx^\mu dx^\nu
+ {\rm e}^{4\phi/3}\, (dy+A_\mu dx^\mu)^2
\;,
\label{KK10}
\end{equation}
in terms of the ten-dimensional metric $g_{\mu\nu}$, (with indices $\mu=0, \dots, 9,$ labelling the coordinates of the ten-dimensional spacetime), 
a dilaton~$\phi$, and a Kaluza-Klein vector $A_\mu$.
Similarly, the components of the eleven-dimensional 3-form are parametrized as
\begin{eqnarray}
   C_{\mu\nu 10} = B_{\mu\nu}    \;, \qquad
   C_{\mu\nu\rho}  = A_{\mu\nu\rho}+3 \,A_{[\mu}\,B_{\nu\rho]}\;,
   \label{BC10}
\end{eqnarray}
in terms of a 2-form $B_{\mu\nu}$ and a 3-form $A_{\mu\nu\rho}$. 
Under the eleven-dimensional gauge transformations (\ref{varC3}) these forms transform as
\begin{eqnarray}
   \delta B_{\mu\nu}  =  2\,\partial_{[\mu} \Lambda_{\nu]}\;, \qquad
   \delta A_{\mu\nu\rho}  = 3\,\partial_{[\mu} \Lambda_{\nu\rho]}-6 \,A_{[\mu}\,\partial_{\nu} \Lambda_{\rho]}\;,
   \label{gauge10}
\end{eqnarray}
respectively, where we denote $\Lambda_\mu=\Lambda_{\mu10}$\,. Accordingly, their gauge invariant field strengths 
are defined as
\begin{eqnarray}
H_{\mu\nu\rho} =  3\,\partial_{[\mu}B_{\nu\rho]}\;, \qquad
   F_{\mu\nu\rho\sigma}  = 4\, \partial_{[\mu}A_{\nu\rho\sigma]}+6\, F_{[\mu\nu} \,B_{\rho\sigma]}
   \;,
\end{eqnarray}
where $F_{\mu\nu}=2\,\partial_{[\mu} A_{\nu]}$ denotes the abelian field strength of the Kaluza-Klein vector from (\ref{KK10}).
Plugging the reduction ansatz (\ref{d11}), (\ref{KK10}), (\ref{BC10}), into the supergravity Lagrangian (\ref{LD11}),
yields the ten-dimensional Lagrangian of type IIA supergravity
\begin{eqnarray}
{\cal L}_{\rm IIA} &=& 
 |e|\, R_{(10)}
  -\frac{1}{4}\, |e|\,{\rm e}^{3\phi/2}\, F_{\mu\nu} F^{\mu\nu} 
   -\frac{1}{2}\, |e|\, \partial_{\mu}\phi\,\partial^{\mu}\phi
\nonumber\\
&&{}
-\frac{1}{48}\,|e|\,{\rm e}^{\phi/2} \, F_{\mu\nu\rho\sigma}F^{\mu\nu\rho\sigma}
-\frac1{12}\, |e|\,{\rm e}^{-\phi}\,H_{\mu\nu\rho} H^{\mu\nu\rho}
\nonumber\\
&&{}
+\frac{1}{144^2} \, \varepsilon^{\mu_0\mu_1\dots\mu_{9}} 
\,F_{\mu_0\mu_1\mu_2\mu_3} \left(3\,F_{\mu_4\mu_5\mu_6\mu_7} \,B_{\mu_8\mu_{9}} 
-8\,H_{\mu_4\mu_5\mu_6} \,A_{\mu_7\mu_{8}\mu_{9}}\right)
\nonumber\\[2ex]
&&{}
+~~{\rm fermions}
\;.
\label{LD10A}
\end{eqnarray}
Here, $|e|$ and $R_{(10)}$ are the determinant of the ten-dimensional vielbein, and the ten-dimensional
Ricci scalar, respectively.
The topological term is invariant under gauge transformations (\ref{gauge10}) up to a total derivative.
The field content of (\ref{LD10A}) matches the IIA supergravity multiplet of (\ref{D10AB}). In particular, 
reduction of the 11D gravitino gives rise to two ten-dimensional gravitini of opposite chirality.
The IIA supergravity Lagrangian may be equivalently expressed in terms of different fundamental 
fields (e.g., using the original components $C_{\mu\nu\rho}$ rather than the $A_{\mu\nu\rho}$ of (\ref{BC10})).
However, all formulations share the property that the field strength $F_{\mu\nu\rho\sigma}$ building the kinetic
term for the 3-form, satisfies a non-trivial Bianchi identity
\begin{equation}
5\,\partial_{[\mu} F_{\nu\rho\sigma\tau]} = 10\,F_{[\mu\nu}H_{\rho\sigma\tau]}
\;.
\end{equation}
Let us also note that the Lagrangian (\ref{LD10A}) admits a 1-parameter massive 
deformation upon deforming the field strengths
\begin{equation}
F_{\mu\nu} \rightarrow F_{\mu\nu} + m\,B_{\mu\nu}
\;,\qquad
F_{\mu\nu\rho\sigma}  \rightarrow
 F_{\mu\nu\rho\sigma} 
 +3\,m\, B_{[\mu\nu} \,B_{\rho\sigma]} 
\;,
\label{mIIA}
\end{equation}
thereby inducing a mass term for the 2-form $B_{\mu\nu}$~\cite{Romans:1985tz}. The resulting theory has an
equivalent description in terms of a 9-form gauge potential which reflects the presence of 
D8-branes in IIA string theory~\cite{Polchinski:1995mt,Bergshoeff:1996ui}.

\subsection{IIB supergravity}
\label{subsec:IIB}

As shown in equation (\ref{D10AB}) above, the bosonic field content of the ten-dimensional type IIB supergravity comprises
the metric, two scalar fields, two 2-form gauge potentials $C_{\mu\nu}{}^\alpha$, $\alpha=1, 2$,
and a selfdual 4-form potential $C_{\mu\nu\rho\sigma}$. Specifically, the latter satisfies a first-order
selfduality equations
\begin{eqnarray}
{F}_{{\mu}{\nu}{\rho}\sigma\tau}&=&
\frac1{5!}\,|e|\,\varepsilon_{{\mu}{\nu}{\rho}\sigma\tau
\mu_1\mu_2\mu_3\mu_4\mu_5}\,
{F}^{\mu_1\mu_2\mu_3\mu_4\mu_5}
\;,
\label{FF55}
\end{eqnarray}
in terms of the field strength
 \begin{equation}
   {F}_{{\mu}_1\ldots{\mu}_5} \ = \ 5\,\partial_{[{\mu}_1}{C}_{{\mu}_2\ldots {\mu}_5]}
  -\frac54 \, 
  \varepsilon_{\alpha\beta}\,{C}_{[{\mu}_1{\mu}_2}{}^{\alpha}{F}_{{\mu}_3{\mu}_4{\mu}_5]}{}^{\beta}\;.
 \label{F5}
 \end{equation}
Here, $ \varepsilon_{\alpha\beta}$ is the antisymmetric tensor in two indices, and 
\begin{equation}
{F}_{\mu\nu\rho}{}^{\alpha}=3\,\partial_{[\mu} C_{\nu\rho]}{}^\alpha
\;,
\end{equation}
 is the abelian field strength for the doublet of 2-forms.
IIB supergravity has been constructed in~\cite{Schwarz:1983wa,Schwarz:1983qr,Howe:1983sra}.
Selfduality equations such as (\ref{FF55}) cannot be derived from a standard action principle. 
As a consequence, these equations are often imposed separately, while the
remaining field equations of IIB supergravity are most conveniently derived from 
a so-called pseudo-action with Lagrangian given by
\begin{equation}
\begin{split}
{\cal L}_{\rm IIB} \ = \ &   |e|\,{R}_{(10)}
+\frac{1}{4}\, |e|\, \partial_{{\mu}}m_{\alpha \beta}\partial^{{\mu}}m^{\alpha \beta}
-\frac{1}{12}\,|e|\,{F}_{{\mu}_1{\mu}_2{\mu}_3}{}^{\alpha}{F}^{{\mu}_1{\mu}_2{\mu}_3}{}^{\beta}m_{\alpha \beta}\\
&-\frac{1}{30} \,|e|\,{F}_{{\mu}_1{\mu}_2{\mu}_3{\mu}_4{\mu}_5}{F}^{{\mu}_1{\mu}_2{\mu}_3{\mu}_4{\mu}_5}\\
&-\frac{1}{864}\,
\varepsilon_{\alpha\beta}\,\varepsilon^{{\mu}_1\ldots {\mu}_{10}}C_{{\mu}_1{\mu}_2{\mu}_3{\mu}_4}{F}_{{\mu}_6{\mu}_7{\mu}_8}{}^{\alpha}{F}_{{\mu}_8{\mu}_9{\mu}_{10}}{}^{\beta}
\\[1ex]
&
+~~{\rm fermions}
\;.
\end{split}
\label{LD10B}
\end{equation}
Here the matrix $m_{\alpha\beta}$ denotes a symmetric $2\times2$ matrix of unit determinant
which is parametrized by the two scalars, $\phi$, $C_0$, of the theory
\begin{equation}
m_{\alpha\beta} = \frac1{\Im\tau}\,
\left(
\begin{array}{cc}
|\tau|^2 &- \Re\tau\\
-\Re\tau & 1
\end{array}
\right)_{\!\!\alpha\beta}\;\,,\qquad
{\tau}=C_0+i \,{\rm e}^{-\phi}
\;.
\label{MSL2}
\end{equation}
Its inverse is denoted as
$m^{\alpha\beta}=\varepsilon^{\alpha\gamma}\varepsilon^{\beta\delta}m_{\gamma\delta}$, 
such that the kinetic term in (\ref{LD10B}) describes the sigma model on the coset space ${\rm SL}(2)/{\rm SO}(2)$, c.f.\ section~\ref{sec:coset},
\begin{equation}
\frac{1}{4}\, \partial_{{\mu}}m_{\alpha \beta}\partial^{{\mu}}m^{\alpha \beta} = -\frac12\,\partial_\mu\phi\partial^\mu\phi
-\frac12\,{\rm e}^{2\phi}\,\partial_\mu C_0 \partial^\mu C_0
\;.
\label{cosetIIB}
\end{equation}
Type IIB supergravity is manifestly invariant under a global ${\rm SL}(2)$
symmetry, acting on all indices $\alpha, \beta$,
inducing a non-linear action on the scalars $\phi$, $C_0$ via (\ref{MSL2}).
Furthermore, the Lagrangian (\ref{LD10B}) is invariant under gauge transformations
\begin{equation}
\begin{split}
\delta {C}_{{\mu} {\nu}}{}^{\alpha}&\ = \ 2\,\partial_{[{\mu}}{\Lambda}_{{\nu}]}{}^{\alpha}\;,\\
\delta {C}_{{\mu}{\nu}{\rho}{\sigma}}& \ = \ 4\,\partial_{[{\mu}}{\Lambda}_{{\nu}{\rho}{\sigma}]}+\frac12\, \varepsilon_{\alpha \beta} {\Lambda}_{[{\mu}}{}^{\alpha}{F}_{{\nu}{\rho}{\sigma}]}{}^{\beta}
\;, 
\end{split}
\label{Ctensor}
\end{equation}
up to total derivatives from variation of the topological term.
It is straightforward to verify that the integrability conditions of the selfduality equations (\ref{FF55}) 
coincide with the second-order field equations obtained by variation of (\ref{LD10B}).
Various alternative action principles for IIB supergravity have been put forward in the literature
in order to also derive the selfduality equations (\ref{FF55}) from a variational principle,
typically at the expense of introducing additional fields
and/or sacrificing manifest (9+1)-dimensional Lorentz 
invariance~\cite{Henneaux:1988gg,Schwarz:1993vs,DallAgata:1997gnw,DallAgata:1998ahf,Hohm:2013vpa,Sen:2015nph,Mkrtchyan:2022xrm}.

The ${\rm SL}(2)$ conventions of (\ref{LD10B}), (\ref{MSL2})
can be translated into the ${\rm SU}(1,1)/{\rm U}(1)$
conventions of \cite{Schwarz:1983qr} by combining the real components of the doublet 
${F}_{\mu\nu\rho}{}^\alpha$ into a complex field strength
\begin{eqnarray}
{F}_{\mu\nu\rho}{} &\equiv &
{F}_{\mu\nu\rho}{}^1+i\,{F}_{\mu\nu\rho}{}^2
\;,
\end{eqnarray}
and parametrizing the matrix $m_{\alpha\beta}$
in terms of a single complex scalar field $B$ as
\begin{eqnarray}
m_{\alpha\beta} &\equiv& (1-BB^*)^{-1} \left(
\begin{array}{cc}
(1-B)(1-B^*) & i\, (B-B^*) \\
 i \,(B-B^*)  & (1+B)(1+B^*)
 \end{array}
\right)_{\alpha\beta}
\;.
\end{eqnarray}
In terms of the complex combinations 
\begin{eqnarray}
G_{\mu\nu\rho}\ \equiv \ (1-BB^*)^{-1/2}\,(F_{\mu\nu\rho}-B\, F_{\mu\nu\rho}^*)\;,\quad
P_{\mu} \ \equiv \ (1-BB^*)^{-1} \partial_{\mu} B\;,
\end{eqnarray} 
the kinetic terms of (\ref{LD10B}) translate into those of \cite{Schwarz:1983qr} with
\begin{eqnarray}
 m_{\alpha\beta} {F}_{{\mu}{\nu}{\rho}}{}^{\alpha}{F}^{{\mu}{\nu}{\rho}\,\beta}
 = G^*_{{\mu}{\nu}{\rho}} G^{{\mu}{\nu}{\rho}}  
\;,\quad
\frac14\,\partial_{{\mu}} m_{\alpha\beta}\partial^{{\mu}} m^{\alpha\beta}
= -2\,P^*_{{\mu}} P^{{\mu}}
\;.
\end{eqnarray}

Let us finally note, that the Lagrangians (\ref{LD10A}), (\ref{LD10B}) of type IIA and type IIB supergravity coincide when truncated to the common field content $\{g_{\mu\nu}, \phi, B_{\mu\nu}=C_{\mu\nu}{}^1\}$, which is the bosonic part of the ${\cal N}=1$ supergravity multiplet from (\ref{D10AB}) --- or the NS-NS sector.

\section{Toroidal reduction}
\label{sec:toroidal}

Maximal supergravities can be obtained by Kaluza-Klein reduction from 
11D and  type IIB supergravity. In particular, the reduction of 
11D supergravity on a $d$-dimensional torus $T^d$  yields the maximal ungauged supergravity in $D=11-d$ dimensions.
More precisely, for $D<10$, this is the unique supergravity theory in $D$ dimensions with 32 real supercharges and no 
fields charged under the abelian gauge group.

In this section, we first discuss the global geometric ${\rm GL}(d)$ symmetry appearing after reduction to $D$ dimensions
as a remnant of the higher-dimensional diffeomorphisms acting on the torus $T^d$. We explicitly perform the dimensional reduction
on a torus, first for pure gravity and next for the $p$-forms, which typically span the matter sector of higher-dimensional supergravities.
In section~\ref{sec:hidden}, we then discuss the enhancement of the geometric symmetry group by the so-called hidden symmetries.

\subsection{Geometric symmetries}
\label{subsec:geometric}

The dimensional reduction of supergravity can be performed most conveniently by using the vielbein 
formalism, see e.g., \cite{Scherk:1979zr}.
We will first consider the reduction of pure gravity in an $(D+d)$-dimensional spacetime
on a $d$-dimensional torus $T^d$ down to $D$ dimensions.
The coordinates of $(D+d)$-dimensional spacetime are split according to
\begin{eqnarray}
x^M\rightarrow \left(x^\mu, y^m\right)\;,\qquad
\mu=0, \dots, D-1\,,\quad m=1, \dots, d\;,
\label{KKsplit}
\end{eqnarray}
and similarly we split the flat Lorentz indices as 
\begin{eqnarray}
A\rightarrow (a,\underline{a})\;,\qquad
a=0, \dots, D-1\,,\quad \underline{a}=1, \dots, d
\;.
\label{KKsplitflat}
\end{eqnarray}
In toroidal dimensional reduction, all fields are taken to be independent of the coordinates $y^m$ of the $d$-torus
\begin{equation}
\partial_m \Phi = 0
\;.
\label{torus_truncation}
\end{equation}
One may think of a normal mode expansion of the fields and drop all modes other than the zero modes.

The local Lorentz invariance in $(D+d)$ dimensions
can be used to bring the vielbein into a triangular form
\begin{equation}
E_M^{\phantom{M}A} = \left( 
\begin{array}{cc}
E_\mu^{\phantom{\mu}a} & E_\mu^{\phantom{\mu}\underline{a}} \\
0&E_m^{\phantom{m}\underline{a}} \\ 
\end{array} \right)\;,
\label{vielbeinD}
\end{equation}      
which breaks the Lorentz group ${\rm SO}(1,D+d-1)$ down to ${\rm SO}(1, D-1)\times {\rm SO}(d)$.
It turns out to be convenient to further parametrize~(\ref{vielbeinD}) as
\begin{eqnarray}
E_M^{\phantom{M}A} &=&
\left(
\begin{array}{cc}
{\rm e}^{\gamma\phi}\,e_\mu{}^a & \,\,{\rm e}^{\phi/d}\,V_m{}^{\underline{a}}\,A_\mu{}^m \\
0 & {\rm e}^{\phi/d}\,V_m{}^{\underline{a}}
\end{array}
\right)
\label{vielbeinreduced}
\;,
\end{eqnarray}
with a matrix $V_m^{\phantom{m}{\underline{a}}}\in{\rm SL}(d)$ 
of unit determinant, such that ${\rm e}^\phi = \det E_m^{\phantom{m}{\underline{a}}}$.
The constant 
\begin{equation}
\gamma =-\frac1{D-2}\;,
\label{gamma}
\end{equation}
is chosen such that plugging~(\ref{vielbeinreduced}) into the $(D+d)$-dimensional
Einstein-Hilbert Lagrangian, one finds
\begin{eqnarray}
{\cal L}^{(D+d)}_{\rm EH} ~=~
|E|\,R_{(D+d)}
&\longrightarrow&
 |e|\, R_{(D)} ~+~ \dots
\;,
\label{LL}
\end{eqnarray}
where $R_{(D)}$ is the Ricci scalar computed from the $D$-dimensional vielbein $e_\mu{}^a$. I.e.,
the reduced theory is directly obtained in the Einstein frame (without a dilaton pre-factor).
The ellipsis in (\ref{LL}) represents the matter couplings in the $D$-dimensional theory,
i.e., the couplings of vector fields $A_\mu{}^m$ and scalar fields $\phi$, $V_m{}^{\underline{a}}$, from (\ref{vielbeinreduced}).
Before working out the explicit form of these terms, it is instructive to analyze the symmetries of the lower-dimensional theory (\ref{LL}) .

The $D$-dimensional theory inherits a number of symmetries
from its higher-dimensional ancestor. Namely, with
the vielbein~(\ref{vielbeinreduced}) 
transforming under infinitesimal diffeomorphisms 
as 
\begin{eqnarray}
\delta_\xi E_M{}^A &=& \xi^N\partial_N  E_M{}^A + E_N{}^A\,\partial_M \xi^N
\;,
\label{trafoE}
\end{eqnarray}
it is straightforward to see that such diffeomorphisms survive in the truncated theory (\ref{torus_truncation})
if the diffeomorphism parameter $\xi^M$ itself satisfies (\ref{torus_truncation}).
In particular, diffeomorphisms of the type $\xi^M=\{\xi^\mu(x),0\}$ generate the $D$-dimensional
diffeomorphisms on the fields $e_\mu{}^a$, $A_\mu{}^n$, $\phi$, and $V_m{}^{\underline{a}}$.
On the other hand, under diffeomorphisms of the type $\xi^M=\{0,\xi^m(x)\}$,
the fields $A_\mu{}^m$ transform as
\begin{equation}
\delta A_{\mu}{}^m ~=~ \partial_\mu \xi^m(x)\;,
\label{gaugeKK}
\end{equation}
whereas the graviton and the scalar fields are left inert.
This shows that the resulting theory is an abelian ${\rm U}(1)^d$ gauge theory
with gauge fields $A_{\mu}{}^m$,
while none of the matter is charged under the gauge group. Accordingly, the
vector fields will couple with a Maxwell-type term
in the reduced theory.

A different type of symmetry can be inferred from internal diffeomorphisms
linear in the compactified coordinates~$y^m$, i.e., of the form 
\begin{equation}
\xi^m(y)=\Lambda^m_{\phantom{m}n}\, y^n
\;.
\label{diffLinY}
\end{equation}
Despite this dependence on the internal coordinates, the action (\ref{trafoE}) of such a diffeomorphism
remains compatible with the truncation~(\ref{torus_truncation}).
Explicitly, this induces a {\rm global} symmetry ${\rm SL}(d)$ acting on the $D$-dimensional matter fields parametrizing (\ref{vielbeinreduced}) as
\begin{eqnarray}
\delta V_m^{\phantom{m}{\underline{a}}} &=& \Lambda^n_{\phantom{n}m} V_n^{\phantom{n}{\underline{a}}} \;,\qquad
\delta A_{\mu}^{\phantom{\mu}m} ~=~ - \Lambda^m_{\phantom{m}n}\, A_{\mu}^{\phantom{\mu}n}
\;,
\label{SLd}
\end{eqnarray}
where we have taken the matrix $\Lambda^m{}_n$ to be traceless $\Lambda^m{}_m\equiv0$.
The trace part of such transformation is slightly more subtle. With
the parametrization~(\ref{vielbeinreduced}), an internal diffeomorphism $\xi^m(y)\propto \,y^m$ also induces a non-trivial action 
on the $D$-dimensional vielbein. It has to be accompanied by the action of the trombone rescaling symmetry of the $(D+d)$-dimensional theory, 
c.f.~(\ref{trombone}), in order to yield a 
proper off-shell symmetry of the $D$-dimensional theory. The combination of 
these transformations induce the action
\begin{eqnarray}
\delta \phi ~=~- \lambda\;,\quad
\delta A_{\mu}^{\phantom{\mu}m} ~=~ 
\lambda\,\beta\,A_{\mu}{}^{m}
\;,
\qquad
\beta=\frac{D+d-2}{d\,(D-2)}
\;,
\label{GL1}
\end{eqnarray} 
with constant $\lambda$, on the $D$-dimensional fields. Together, the transformations (\ref{SLd}) and (\ref{GL1}) generate a global 
${\rm GL}(d)$ symmetry of the $D$-dimensional theory. We will refer to this group as the geometric symmetries
of the theory, as they have their origin in the diffeomorphisms on the internal torus.
Let us stress, that the enhancement from 
${\rm SL}(d)$ to ${\rm GL}(d)$ requires the higher-dimensional scaling symmetry (\ref{trombone})
and is no longer realized in the presence of higher curvature corrections.

Finally, local Lorentz invariance is also a symmetry of the higher-dimensional theory. 
As mentioned above, the upper triangular form (\ref{vielbeinD}) of the vielbein
breaks the original ${\rm SO}(1,D+d-1)$ down to 
${\rm SO}(1,D-1) \times {\rm SO}(d)$,
of which the first factor acts as $D$-dimensional Lorentz transformation on $e_\mu{}^a$
and the second factor acts as an additional {\rm local} symmetry on $V_m^{\phantom{m}\underline{a}}$
\begin{equation}
\delta V_m^{\phantom{m}\underline{a}}= V_m^{\phantom{m}\underline{b}} \Lambda_{\underline{b}}^{\phantom{b}\underline{a}}(x)
\;,\qquad
\Lambda(x)\in\mathfrak{so}(N)
\;.
\label{localH}
\end{equation}    
This shows that not all components of the matrix $V_m^{\phantom{m}\underline{a}}$ correspond to physical scalars
as we make explicit in the following.

\subsection{Reduction of pure gravity}
\label{subsec:reductiongravity}

For pure gravity, the fields of the $D$-dimensional theory are the various components of the higher-dimensional vielbein (\ref{vielbeinreduced}). 
The global and local symmetries of the $D$-dimensional theory, identified in the previous subsection, almost uniquely fix the form of
the resulting two-derivative action. 
Explicitly, plugging the ansatz~(\ref{torus_truncation}), (\ref{vielbeinreduced}), into the $(D+d)$-dimensional
Einstein-Hilbert Lagrangian, yields the completion of (\ref{LL}) 
\begin{eqnarray}
{\cal L}^{(D+d)}_{\rm EH}~=~
|E|\,R_{(D+d)}
&\longrightarrow&
   |e| \, R_{(D)}
-  |e| \, \mbox{Tr} \left[P_\mu P^\mu \right]-\beta\,|e|\,
\partial_\mu \phi \,\partial^\mu \phi
\nonumber\\
&&{} 
-\frac1{4}\,|e|\,{\rm e}^{2\,\beta\,\phi}\, M_{mn}\,
F_{\mu\nu}{}^m\,F^{\mu\nu\,n}
\nonumber\\[.5ex]
&&{} 
+~\mbox{total derivatives}
\;,
\label{reductionG}
\end{eqnarray} 
with $\beta$ from (\ref{GL1}). The vector fields appear with abelian field strengths given by
$F_{\mu\nu}{}^m=2\,\partial_{[\mu}A_{\nu]}{}^m$
and a Maxwell term with the scalar dependent metric
\begin{eqnarray}
M_{mn}= V_m{}^{\underline{a}}V_n{}^{\underline{b}}\,\delta_{\underline{ab}}\;.
\label{Mbeta}
\end{eqnarray}
The scalar fields $V_m^{\phantom{m}\underline{a}}$ describe a coset space sigma model 
\begin{equation}
{\rm G}/{\rm K}={\rm GL}(d)/{\rm SO}(d)\;, 
\label{cosetGLd}
\end{equation}
with the $\mathfrak{sl}(d)$ currents defined by
\begin{eqnarray}
\delta^{\underline{ac}}\,(V^{-1})_{\underline{c}}{}^{m} \partial_\mu V_m{}^{\underline{b}} &\equiv& Q_{\mu}{}^{[\underline{ab}]}+ P_{\mu}{}^{(\underline{ab})}
\;,
\label{SLd_current}
\end{eqnarray}
and decomposed into their antisymmetric and symmetric parts. The target space ${\rm SL}(d)/{\rm SO}(d)$ is given by
\begin{eqnarray}
- \mbox{Tr} \left[P_\mu P^\mu \right]
&=&\frac14\,\partial_\mu M_{mn} \, \partial^\mu (M^{-1})^{mn} 
\;,
\label{trPP}
\end{eqnarray}
with the matrix $M_{mn}$ from (\ref{Mbeta}).
Indeed, this kinetic term is invariant under the local symmetry transformations (\ref{localH}), showing that the matrix $V_m^{\phantom{m}\underline{a}}$
carries 
\begin{equation}
{\rm dim}\,\left({\rm SL}(d)\big/{{\rm SO}(d)}\right) = \frac12\,(d-1)(d+2)
\;,
\end{equation}
physical scalar fields.

Up to its relative coefficients, the Lagrangian (\ref{reductionG}) is the unique
two-derivative Lagrangian for this field content, which is compatible with the global symmetries (\ref{SLd}), (\ref{GL1}),
as well as with the gauge symmetries (\ref{gaugeKK}), (\ref{localH}), whose presence was deduced from the
higher-dimensional diffeomorphism and Lorentz symmetries.

\subsection{Reduction of $p$-forms}
\label{subsec:reductionsupergravity}

The bosonic sector of higher-dimensional supergravities typically combines Einstein gravity
with $p$-form matter couplings, such as the 3-form couplings (\ref{112}) of 11D supergravity.
Upon dimensional reduction on a torus $T^d$, this matter sector gives rise to additional fields, couplings,
and symmetries in the lower-dimensional theory.
Before performing the explicit toroidal reduction of the Lagrangian (\ref{112}), it is instructive to first study 
the behavior of the additional fields
w.r.t.\ the global ${\rm GL}(d)$ symmetry (\ref{SLd}), (\ref{GL1}).

Although we will mostly be interested in the reduction of a 3-form from 11 dimensions, the analysis
is straightforward in general spacetime dimensions. Let us consider a $p$-form ${C}_{ M_1  \cdots M_p}$
in $(D+d)$ spacetime dimensions,
such that its various components
give rise to $D$-dimensional $p$, $p-1$, \dots, $(p-d)$-forms.
The precise reduction ansatz corresponds to a split in the flat basis (\ref{KKsplitflat}),
i.e., the $D$-dimensional $k$-forms are built as
\begin{eqnarray}
A_{\mu_{1} \cdots \mu_{k}\,m_{k+1} \cdots m_p} &=&
P_{\mu_{1}}{}^{M_{1}} \cdots P_{\mu_{k}}{}^{M_{k}}\,
C_{ M_{1} \cdots M_{k}\,m_{k+1} \cdots m_p}
\;,
\label{embedA}
\end{eqnarray}
with $P_\mu{}^M \equiv \{ \delta_\mu{}^\nu, -A_\mu{}^m \}$\,. 
This ansatz (\ref{embedA}) is such that the lower-dimensional fields remain invariant
under the Kaluza-Klein gauge transformations (\ref{gaugeKK}).
The transformation behavior of these fields under the 
global ${\rm SL}(d)$ symmetry from (\ref{SLd}) can be computed from the action of diffeomorphisms (\ref{diffLinY})
and follows from their index structure in the internal indices $m_1$, $m_2$, \dots, e.g.,
\begin{equation}
\delta A_{\mu\,mn} ~=~  \Lambda^k{}_m\, A_{\mu\,kn}+\Lambda^k{}_n\, A_{\mu\,mk}
\;,\qquad
\mbox{etc..}
\end{equation}
Their charge under the  ${\rm GL}(1)$ symmetry from (\ref{GL1}) is slightly more tedious to determine,
since it involves the $(D+d)$-dimensional trombone symmetry as discussed above.
One obtains (see e.g., \cite{deWit:2002vt})
\begin{eqnarray}
\delta_\lambda A_{\mu_{1} \cdots \mu_{k}\,m_{k+1} \cdots m_p}
&=& 
-\lambda\,\Big(p\,\gamma +(p-k)\,\beta\Big)\,
A_{\mu_{1} \cdots \mu_{k}\,m_{k+1} \cdots m_p} \;,
\label{actionGL1}
\end{eqnarray}
with $\gamma$ and $\beta$ from (\ref{gamma}), (\ref{GL1}).
The higher-dimensional tensor gauge symmetries
\begin{equation}
\delta {C}_{ M_1  \cdots M_p} = p\,\partial_{[M_1} \Lambda_{M_2 \cdots M_p]}
\;,
\label{tensorP}
\end{equation}
give rise to the lower-dimensional gauge symmetries of the $k$-forms.
Due to the reduction ansatz (\ref{embedA}), these symmetries in general mix forms of different degree 
with a non-linear action.

For a sufficiently large torus, i.e., for $d\ge p$ the reduction (\ref{embedA})
adds $d\choose p$ scalar fields $A_{m_1 \dots m_p}$ to the scalar sector of
the $D$-dimensional theory. In analogy to (\ref{diffLinY}), the
higher-dimensional tensor gauge transformations linear in
the compactified coordinates
\begin{equation}
\Lambda_{m_1\dots m_{p-1}}(y)=\xi_{m_1 \dots m_p}\, y^{m_p}\;,
\label{shiftY}
\end{equation} 
induce additional global shift symmetries
\begin{eqnarray}
\delta_\xi\, A_{m_1 \dots m_p} &=& 
\xi_{m_1 \dots m_p}
\;,
\label{shift1}
\end{eqnarray}
on these scalar fields. These symmetries enhance the global $\mathfrak{gl}(d)$ from   (\ref{SLd}), (\ref{GL1}), to a non-semisimple algebra 
of the type
\begin{equation}
\mathfrak{g}_{\rm nss} = \mathfrak{g}_0  \oplus \mathfrak{n}_+
\;,
\label{gnss}
\end{equation}
where $\mathfrak{g}_0$ combines the geometric $\mathfrak{gl}(d)$ with other potential global symmetries 
of the higher-dimensional theory, while the nilpotent $\mathfrak{n}_+$ combines all the shifts of type (\ref{shift1}).
For example, for reductions from 11D supergravity,  $\mathfrak{g}_0=\mathfrak{gl}(d)$, whereas 
for reductions from IIB supergravity $\mathfrak{g}_0=\mathfrak{gl}(d)\oplus\mathfrak{sl}(2)$.
The algebra (\ref{gnss}) is graded w.r.t.\ the $\mathfrak{gl}(1)\subset\mathfrak{gl}(d)$ of (\ref{GL1}), under which $\mathfrak{g}_0$ are the zero modes.
Moreover, it follows from (\ref{actionGL1}) that all generators of $\mathfrak{n}_+$ have positive charge under $\mathfrak{gl}(1)$.
We emphasize once more, that all the global symmetries (\ref{gnss}) of the lower-dimensional theory have a direct geometrical
origin by the higher-dimensional local diffeomorphism and tensor gauge symmetries.

Let us also note that part of the shift symmetries (\ref{shift1}) may arise from the dual higher-dimensional $p$-forms. For example, we have noted in
(\ref{F47}) the existence of the dual 6-form in 11D supergravity. Upon toroidal reduction on a sufficiently large torus $T^d$ 
with $d\ge6$, the associated gauge symmetries induce shift symmetries (\ref{shift1}) on the scalar fields descending from the 6-form. 
This indicates that the full symmetry algebra (\ref{gnss}) in general is only visible after taking into account all the fields together with their duals.

Furthermore, the symmetry induced by the gauge transformations (\ref{shiftY}) does not only act on scalar fields via the shift 
(\ref{shift1}) but may in general also 
have a non-trivial action on some of the $p$-forms. Consider the gauge transformations (\ref{shiftY})
\begin{equation}
\Lambda_{m_1m_2}(y)=\xi_{m_1m_2m_3}\, y^{m_3}\;,
\label{shiftY3}
\end{equation} 
in 11D supergravity. While they induce the shifts (\ref{shift1}) on the scalar fields $A_{m_1m_2m_3}$, they
also have a nontrivial action on the $p$-forms descending from the dual 6-form according to its gauge transformation~(\ref{varC6}).
For example,  the $D$-dimensional 3-forms $C_{\mu\nu\rho\, m_1m_2m_3}$, transform as
\begin{equation}
\delta C_{\mu\nu\rho\, m_1m_2m_3} =
 - \frac32\,  C_{\mu\nu\rho}\,\xi_{m_1m_2m_3} 
 \;,
 \label{c6c3}
\end{equation}
and similar for the lower-rank forms.

As an illustration for the reduction of $p$-forms, let us
perform the explicit reduction of the 3-form Lagrangian (\ref{112}) of 11D supergravity. The reduction 
ansatz (\ref{embedA}) identifies the $D$-dimensional scalars, 1-forms, 2-forms, and 3-form as 
 \begin{equation}
  \begin{split}
   A_{mnk} \ &= \ C_{mnk}\;, \\
   A_{\mu\,mn} \ &= \ C_{\mu mn}-A_{\mu}{}^k\,C_{kmn}\;, \\
   A_{\mu\nu\,m} \ &= \ C_{\mu\nu m}-2\, A_{[\mu}{}^n\,C_{\nu]mn}+A_{\mu}{}^n A_{\nu}{}^k\,C_{mnk}\;, \\
   A_{\mu\nu\rho} \ &= \ C_{\mu\nu\rho}-3\, A_{[\mu}{}^m\,C_{\nu\rho]m}+3\, A_{[\mu}{}^m A_{\nu}{}^n\, C_{\rho]mn}
   -A_{\mu}{}^m A_{\nu}{}^n A_{\rho}{}^k\,C_{mnk}\;,
  \end{split}
  \label{comp3form}
 \end{equation}  
 in terms of the components of $C_{KLM}$.
 The $y$-independent eleven-dimensional gauge transformations (\ref{tensorP}) translate into
   \begin{equation}\label{gaugevar}
  \begin{split}
   \delta A_{\mu mn} \ &= \ \partial_{\mu}\Lambda_{mn}\;, \\
   \delta A_{\mu\nu\, m} \ &= \ 2 \,\partial_{[\mu}\Lambda_{\nu]m}-F_{\mu\nu}{}^n\Lambda_{mn}\;, \\
   \delta A_{\mu\nu\rho} \ &= \ 3\,\partial_{[\mu}\Lambda_{\nu\rho]}-3\, F_{[\mu\nu}{}^m \Lambda_{\rho]m}\;, 
  \end{split}
 \end{equation}  
where the lower-dimensional gauge parameters have been embedded into $\Lambda_{MN}$ in analogy to (\ref{embedA}).
Likewise, the gauge invariant field strengths in $D$ dimensions are defined via
 \begin{eqnarray}
F_{\mu_{1} \cdots \mu_{k}\,m_{k+1} \cdots m_4} &=&
4\,P_{\mu_{1}}{}^{M_{1}} \cdots P_{\mu_{k}}{}^{M_{k}}\,
\partial_{ [M_{1}} C_{ \cdots M_{k}\,m_{k+1} \cdots m_4]}
\;,
\label{embedF}
\end{eqnarray}
and take the explicit form
  \begin{equation}\label{invfieldstr}
  \begin{split}
    F_{\mu\, nkl} \ &= \ \partial_{\mu}A_{nkl}\;, \\
   F_{\mu\nu\, mn} \ &= \ 2 \partial_{[\mu}A_{\nu]mn}+F_{\mu\nu}{}^k A_{kmn}\;, \\
   F_{\mu\nu\rho\, m} \ &= \  3\partial_{[\mu}A_{\nu\rho]m}+3 F_{[\mu\nu}{}^n A_{\rho]mn}
 \;, \\
   F_{\mu\nu\rho\sigma} \ &= \ 4 \partial_{[\mu}A_{\nu\rho\sigma]}+6 F_{[\mu\nu}{}^m A_{\rho\sigma]m}\;.  
  \end{split}
 \end{equation}   
They satisfy non-standard non-linear Bianchi identities
  \begin{equation}\label{Bianchi123}
  \begin{split}
   3 \,\partial_{[\mu} F_{\nu\rho]mn} \ &= \ 3\, F_{[\mu\nu}{}^k F_{\rho]kmn}\;, \\
   4 \,\partial_{[\mu}F_{\nu\rho\sigma]m} \ &= \ 6\, F_{[\mu\nu}{}^n F_{\rho\sigma]mn}\;, \\
   5 \,\partial_{[\mu}F_{\nu\rho\sigma\lambda]} \ &= \ 10\, F_{[\mu\nu}{}^m F_{\rho\sigma\lambda]m}\;. 
  \end{split}
 \end{equation}    
Putting everything together, the kinetic term for the 3-form (\ref{112})  
reduces to
  \begin{equation}
  \label{reductionC}
 \begin{split}
 {\cal L}_{\text{\,kin}} 
 \ &= \  -\frac{1}{48}|E|\,
 F^{KLMN}F_{KLMN} \\
   \ &= \ -\frac{1}{48}\,|e|\,\Big(
   {\rm e}^{-6\gamma\phi}\, F^{\mu\nu\rho\sigma}F_{\mu\nu\rho\sigma}
   +4\,{\rm e}^{-(6\gamma+2\beta)\phi}\, M^{mn}F^{\mu\nu\rho}{}_{m} F_{\mu\nu\rho\,n}
  \\ &\qquad\qquad \qquad 
  +6\,{\rm e}^{-(6\gamma+4\beta)\phi}\, M^{mn}M^{kl}F^{\mu\nu}{}_{mk} F_{\mu\nu\, nl}
  \\ &\qquad\qquad \qquad 
  +4\,{\rm e}^{-(6\gamma+6\beta)\phi}\,M^{mn}M^{kl}M^{pq} \partial^{\mu}A_{mkp} \partial_{\mu}A_{nlq}
 \Big)
   \;,
\end{split}  
\end{equation} 
with $\gamma$ and $\beta$ from (\ref{gamma}) and (\ref{GL1}),
and the scalar dependent matrix $M^{mn}$ denoting the inverse of (\ref{Mbeta}).
The reduced Lagrangian provides the kinetic terms for the lower-dimensional forms. It is straightforward to check
that the dilaton powers precisely ensure invariance of the action under the GL(1) scaling symmetry
(\ref{GL1}), (\ref{actionGL1}). Also the invariance under constant shifts (\ref{shift1}) is manifest.

Finally, reduction of the Chern-Simons term in (\ref{112}) gives rise to a lengthy topological term in $D$
dimensions. It is most compactly described by writing the original Chern-Simons term as the boundary contribution
of some twelve-dimensional integral of
\begin{equation}
d{\cal L}_{\rm top} = F^{(4)} \wedge F^{(4)} \wedge F^{(4)} \;,
\label{dCS}
\end{equation}
and to reduce the r.h.s.\ of this equation in terms of the different components (\ref{invfieldstr}).

The straightforward toroidal reduction of 11D supergravity thus gives a lower-dimensional
theory with a global symmetry group of the type (\ref{gnss}) with semisimple part $\mathfrak{g}_0=\mathfrak{gl}(d)$.
Before we discuss the further enhancement of the global symmetry group by hidden symmetries
in the next section, let us 
briefly spell out the case $d=2$, i.e., the reduction to $D=9$ maximal supergravity.

\subsection{Maximal $D=9$ supergravity}
\label{subsec:maximalD9}

We first consider the case $d=2$, i.e., the reduction
of 11D supergravity on a two-torus $T^2$. 
From (\ref{gamma}) and (\ref{GL1}), we find the values $\beta=\frac9{14}$, $\gamma=-\frac17$.
From the general structure given in the previous sections, we read off the resulting Lagrangian in $D=9$ dimensions as
the sum of (\ref{reductionG}), (\ref{reductionC}),
and a nine-dimensional CS term as
\begin{eqnarray}
{\cal L}^{(9)}_{\rm EH}
&=&
   |e| \, R_{(9)}
-  |e| \, \mbox{Tr} \left[P_\mu P^\mu \right]-\frac72\,|e|\,
\partial_\mu \phi \,\partial^\mu \phi
\nonumber\\
&&{} 
-\frac1{4}\,|e|\,{\rm e}^{3\,\phi}\, M_{mn}\,
F_{\mu\nu}{}^m\,F^{\mu\nu\,n}
   -\frac{1}{8}\,|e|\,\,{\rm e}^{-4\phi}\, F^{\mu\nu}{} F_{\mu\nu}
\nonumber\\
&&{}
    -\frac{1}{12}\,|e|\,\,{\rm e}^{-\phi}\, M^{mn}F^{\mu\nu\rho}{}_{m} F_{\mu\nu\rho\,n}
 -\frac{1}{48}\,|e|\,   {\rm e}^{2\phi}\, F^{\mu\nu\rho\sigma}F_{\mu\nu\rho\sigma}
\nonumber\\[1ex]
&&{}
+{\cal L}_{\rm top}
   \;,
   \label{9D}
\end{eqnarray} 
where for convenience we have rescaled the dilaton as $\phi\rightarrow\frac73\,\phi$
and furthermore set
\begin{equation}
 A_{\mu m n} = A_\mu\,\varepsilon_{mn}
\;,\quad
 F_{\mu\nu m n} = F_{\mu\nu}\,\varepsilon_{mn}
 \;.
\end{equation}
As discussed above, the global symmetry group of this theory is given by
\begin{equation}
{\rm GL}(2)  = {\rm SL}(2) \times {\rm GL}(1)
\;,
\label{GL2GL1}
\end{equation}
and there is no further symmetry enhancement, as shift symmetries of the type (\ref{shift1}) are absent.

The remarkable property of the theory (\ref{9D}) is the fact that the very same theory is obtained by
reducing ten-dimensional IIB supergravity on $S^1$. This is consistent with the fact that only a single
maximal supermultiplet exists in $D=9$ dimensions.
For the IIB reduction, the origin of the global symmetry (\ref{GL2GL1}) is the geometric ${\rm GL}(1)$
of the circle, together with the ${\rm SL}(2)$ symmetry of the IIB theory.
The fields from (\ref{9D}) have different higher-dimensional origin according to the scheme discussed
in sections~\ref{subsec:reductiongravity}, and \ref{subsec:reductionsupergravity} above, specifically
\begin{eqnarray}
{11{\rm D}} &:&
\mbox{metric:}\,\{ g_{\mu\nu}, A_\mu{}^m, \phi, M_{mn} \}\;,\nonumber\\
&&\mbox{3-form:} \,\{ A_\mu, A_{\mu\nu\,m}, A_{\mu\nu\rho} \}\;,
\nonumber\\[1ex]
\mbox{IIB} &:& 
\mbox{metric:} \,\{g_{\mu\nu}, A_\mu, \phi \}\;,
\quad
\mbox{scalars:}\,\{M_{mn} \}\;,\nonumber\\
&&\mbox{2-form:}\,\{A_\mu{}^m, A_{\mu\nu\,m} \}\;,
\quad
\mbox{4-form:}\,\{A_{\mu\nu\rho} \}
\;.
\label{9-10-11}
\end{eqnarray}

The presence of Chern-Simons terms in (\ref{LD11}) and (\ref{LD10B}) is indispensable for the equivalence of
the two theories after toroidal reduction. It gives rise to non-trivial Bianchi identities for dual fields (\ref{Bianchi7}) akin to those
appearing after dimensional reduction (\ref{Bianchi123}), thus allowing for the identification of fields of different higher-dimensional origin (\ref{9-10-11}).
A detailed discussion of $D=9$ supergravity and its higher-dimensional embeddings is given in~\cite{Bergshoeff:2002nv}.

\section{Hidden symmetries}
\label{sec:hidden}

We have seen in the previous section that toroidal reduction of higher-dimensional
supergravity theories induces lower-dimensional supergravity theories with manifest
global symmetries descending from higher-dimensional diffeomorphism and
tensor gauge symmetries, spanning an algebra of the type (\ref{gnss}).
It is one of the most remarkable facts about these theories that on top of the
geometric symmetries (\ref{gnss}), the lower-dimensional supergravities possess further
so-called hidden global symmetries which only become apparent after toroidal reduction
and proper redefinition of the fields~\cite{Cremmer:1979up,Julia:1980gr,Julia:1982gx,Julia:1982tm}.

As part of the general pattern, dubbed the `silver rules of supergravity' \cite{Julia:1997cy}, the full
algebra of global symmetries of the lower-dimensional supergravity is given by the extension of (\ref{gnss})
into a semisimple algebra
\begin{equation}
\mathfrak{g} = \mathfrak{n}_- \oplus \mathfrak{g}_0  \oplus \mathfrak{n}_+
\;.
\label{gss}
\end{equation}
The hidden symmetries combine into a nilpotent algebra $\mathfrak{n}_-$, which completes (\ref{gnss}) into a semisimple algebra.
W.r.t.\ $\mathfrak{gl}(d)$, the generators of $\mathfrak{n}_-$ transform in the 
representation dual to the generators of $\mathfrak{n}_+$. In particular, they carry 
negative charge under 
$\mathfrak{gl}(1)\subset\mathfrak{gl}(d)$.
The fact that the dimension and structure of the algebra of hidden symmetries precisely fits the expansion (\ref{gss})
depends of course strongly on the field content and the couplings of the higher-dimensional 
supergravity theory. For generic couplings, no such symmetry enhancement would occur. 
Already a different pre-factor in front of the Chern-Simons term (\ref{112}) of 11D
supergravity would prevent the symmetry enhancement in all lower-dimensional theories.
This is where supersymmetry comes to play its role, although here we only focus on the bosonic sectors of theory.
For example, the coefficients in (\ref{112}) allowing the symmetry enhancement (\ref{gss}) are precisely the coefficients 
that were fixed by supersymmetry of the 11D action.

The global symmetry algebra (\ref{gss}) acts on all fields of the theory. While on the $p$-forms
its action is necessarily linear as imposed by compatibility with gauge symmetry, 
the action of the hidden symmetries on the scalar fields is in general non-linear.
It is most elegantly described by the isometries of the coset space
\begin{equation}
{\rm G}/{\rm K}
\;,
\end{equation}
with ${\rm G}={\rm Lie}\,\mathfrak{g}$, and ${\rm  K}$ its maximal compact subgroup. 
For maximal supergravity, the resulting global symmetry groups ${\rm G}$ build the 
${\rm E}_{d(d)}$ series of non-compact exceptional Lie groups in the Dynkin classification,
with Dynkin diagram given in Figure~\ref{fig:dynkin} above.\footnote{The subscripts in parentheses in
${\rm E}_{d(d)}$ specify the particular real
form of the exceptional groups. Specifically, it denotes the difference between non-compact and compact generators of the associated algebra. 
For maximal supergravity the global symmetry groups always appear in their split form, i.e., the maximally non-compact form of the group.}
For small values of $d$, the series 
degenerates into the classical Lie groups
\begin{equation}
{\rm E}_{5(5)} \simeq {\rm SO}(5,5)\;,\quad
{\rm E}_{4(4)} \simeq {\rm SL}(5)\;,\quad
{\rm E}_{3(3)} \simeq {\rm SL}(3) \times {\rm SL}(2)
\;,
\end{equation}
as can be extrapolated from the general Dynkin diagram of Figure~\ref{fig:dynkin}.
The full set of coset spaces is listed in Table~\ref{table:groups}.

Before going through the various cases,
we first briefly review the structure of such coset spaces and their isometries.

\begin{table}[tb]
\begin{center}
{\small
\begin{tabular}{c||c|}
\;\;$D$ \;\;&${\rm G}/{\rm K}$  
  \\
\hline
9   & ${\rm GL}(2)/{\rm SO}(2)$  \\
8   & $\big({\rm SL}(2)\! \times \!{\rm SL}(3)\big)/\big({\rm SO}(2)\! \times \!{\rm SO}(3)\big)$ \\
7   & ${\rm SL}(5)/{\rm SO}(5)$ \\
6   &  ${\rm SO}(5,5)/\big({\rm SO}(5)\! \times \!{\rm SO}(5)\big)$\\
5   & E$_{6(6)}/{\rm USp}(8)$ \\
4   & E$_{7(7)}/{\rm SU}(8)$ \\
3   & E$_{8(8)}/{\rm SO}(16)$ \\
2   & E$_{9(9)}/{\rm K}({\rm E}_9)$
\end{tabular}
}
\end{center}
\caption{\small
Global symmetry groups ${\rm G}$ and their compact subgroups ${\rm K}$
in maximal supergravity in the various spacetime dimensions. 
For $D=2$, the group ${\rm E}_{9(9)}$ is the
(centrally extended) affine extension of the group
${\rm E}_{8(8)}$,
${\rm K}({\rm E}_9)$ denotes its maximal compact subgroup.
}\label{table:groups}
\end{table}

\subsection{Coset spaces}
\label{sec:coset}

After toroidal reduction, the scalar fields of maximal supergravity theories are most conveniently 
described by a coset space sigma model.
We have already encountered this structure in section~\ref{subsec:reductiongravity} in the reduction of pure gravity on a torus $T^d$
with the scalars parametrizing the target space ${\rm GL}(d)/{\rm SL}(d)$ according to (\ref{trPP}).
Including the higher-dimensional $p$-forms, this space gets enhanced by the additional scalar fields into
a larger coset space
\begin{equation}
 {\rm GL}(d)/{\rm SL}(d)  \xhookrightarrow{}  {\rm G}/{\rm K}
\;.
\label{coset}
\end{equation}
Here, ${\rm G}$ is the Lie group associated to the algebra $\mathfrak{g}$ in (\ref{gss}), and ${\rm  K}$ denotes its maximal compact subgroup.
The coset space is described by a representative (or vielbein) ${\cal V}\in {\rm G}$ with local gauge invariance
\begin{equation}
\delta {\cal V} = {\cal V} k(x)\;,\qquad
k(x) \in \mathfrak{k}
\;,
\label{coset_gauge}
\end{equation}
with ${\rm K}={\rm Lie}\,\mathfrak{k}$\,.
In analogy to (\ref{SLd_current}), (\ref{trPP}), the Lagrangian for the coset space sigma model 
is built by decomposing the left invariant scalar currents as
\begin{equation}
{\cal V}^{-1}\partial_\mu {\cal V} = Q_\mu + P_\mu
\;,
\label{QP}
\end{equation}
with $Q_\mu\in\mathfrak{k}$ and $P_\mu\in\mathfrak{p}$, according to the 
orthogonal decomposition
\begin{equation}
\mathfrak{g} = \mathfrak{k} \oplus \mathfrak{p}
\;.
\end{equation}
The Lagrangian 
\begin{equation}
{\cal L}_{\rm coset} = -\frac12\,|e|\,{\rm Tr}\,(P_\mu P^\mu)
\;,
\label{LPP}
\end{equation}
then is invariant under the gauge transformations (\ref{coset_gauge}), under which
\begin{equation}
\delta P_\mu = [ P_\mu , k(x) ]
\;.
\end{equation}
The gauge symmetry (\ref{coset_gauge}) can be fixed by imposing a sufficient number of conditions 
on the vielbein ${\cal V}$, such that it is uniquely parametrized by 
\begin{equation}
n={\rm dim}\,{\rm G}-{\rm dim}\,{\rm K}
\;,
\end{equation}
scalar fields, corresponding to a choice of coordinates on the target space (\ref{coset}).
The Lagrangian (\ref{LPP}) remains invariant under the global symmetry
\begin{equation}
\delta_g {\cal V} = g {\cal V} - {\cal V} k_g\;,\qquad
g\in\mathfrak{g}\;,\;\; k_g\in\mathfrak{k}
\;,
\label{actionGK}
\end{equation}
combining left multiplication on ${\cal V}$ with a compensating gauge transformation (\ref{coset_gauge})
in order to restore the fixed gauge. The action (\ref{actionGK}) describes the infinitesimal action of the
isometry group ${\rm G}$ on the $n$ coordinates of the target space (\ref{coset}). 
In particular, it encodes the action of the global symmetry group ${\rm G}$ on the 
fermionic sector of the theory. Before gauge fixing the local symmetry (\ref{coset_gauge}), the fermions of the theory
appear as singlets under ${\rm G}$ but transform under local ${\rm K}$ transformations  (\ref{coset_gauge}).
Derivatives are covariantized with the composite connection $Q_\mu$ from (\ref{QP}). After gauge fixing,
the fermions inherit a non-trivial action of the global symmetry group ${\rm G}$ by means of the compensating
$\mathfrak{k}$-transformation $k_g\in\mathfrak{k}$ of~(\ref{actionGK}).

In the context of toroidal reduction  of supergravity, 
a natural gauge fixing for the vielbein ${\cal V}$ is the triangular gauge, in which
this matrix is put to the form
\begin{equation}
{\cal V}  =  {\rm exp}(\phi^a N_a)\, V_{{\rm G}_0} 
\;,\quad
V_{{\rm G}_0}\in{\rm G}_0={\rm Lie}\,\mathfrak{g}_0 \;,
\label{triangular}
\end{equation}
where the right factor $V_{{\rm G}_0}$ lives in the Lie group associated with the algebra of zero charge 
generators $\mathfrak{g}_0$ in (\ref{gss}) and the $N_a$ denote the generators of the nilpotent algebra $\mathfrak{n}_+$.
The matrix $V_{{\rm G}_0}$ is built from the internal part of the higher-dimensional vielbein 
$V_m{}^{\underline{a}}$ introduced in (\ref{vielbeinreduced})
(together with the other scalars of the higher-dimensional theory), while the scalars $\phi^a$ describe the scalars 
descending from the higher-dimensional $p$-forms.

As a result, the action (\ref{actionGK}) of $\mathfrak{g}_0$ induces a linear action on the scalar fields $\phi^a$,
in accordance with the representation of the associated generators $N_a$. The action (\ref{actionGK}) of $\mathfrak{n}_+$
does not require a compensating gauge transformation, $k_{\mathfrak{n}_+}=0$,
and induces shift symmetries on the scalars $\phi^a$, generating the transformations (\ref{shift1}).
In contrast, the action (\ref{actionGK}) of the hidden symmetries $\mathfrak{n}_-$ induces a compensating gauge
transformation and thereby a non-linear action on the scalar fields.

As an illustration, let us evaluate the transformation (\ref{actionGK}) for the coset space ${\rm SL}(2)/{\rm SO(2)}$
which appears in the matter sector of various supergravity theories.
With the $\mathfrak{sl}(2)$ algebra generators given by
\begin{eqnarray}
{\bf h}=
\left(
\begin{array}{cc}
1&0\\
0&-1
\end{array}
\right)
\;,\qquad
{\bf e}=
\left(
\begin{array}{cc}
0&1\\
0&0
\end{array}
\right)
\;,\qquad
{\bf f}=
\left(
\begin{array}{cc}
0&0\\
1&0
\end{array}
\right)
\;,
\label{hef_gen}
\end{eqnarray}
the decomposition (\ref{gss}) of the algebra corresponds to
\begin{equation}
\mathfrak{sl}(2) = \mathfrak{n}_- \oplus \mathfrak{g}_0  \oplus \mathfrak{n}_+ = 
\langle {\bf f} \rangle \oplus 
\langle {\bf h} \rangle \oplus 
\langle {\bf e} \rangle 
\;.
\label{gss_sl2}
\end{equation}
Accordingly, the matrix ${\cal V}$ in triangular gauge~(\ref{triangular})
can be parametrized as
\begin{eqnarray}
{\cal V} &=& 
{\rm exp}(C\,{\rm\bf e}) \,{\rm exp}(\tfrac12{\phi \,{\rm\bf h}})
=
\left(
\begin{array}{cc}
{\rm e}^{\phi/2} & {\rm e}^{-\phi/2}\,C \\
0 & {\rm e}^{-\phi/2}
\end{array}
\right)
\;.
\end{eqnarray}
The action~(\ref{actionGK}) then induces the transformation 
\begin{eqnarray}
\delta_{\bf h}\,\phi=2
\;,\;\;
\delta_{\bf h}\,C=2C
\;,\quad
\delta_{\bf e}\,C=1
\;,\quad
\delta_{\bf f}\,\phi=-2C
\;,\;\;
\delta_{\bf f}\,C={\rm e}^{2\phi}-C^2
\;,
\label{hef}
\end{eqnarray}
on the scalars $\phi$, $C$.
This shows how the algebra $\mathfrak{g}_0=\langle {\bf h} \rangle$ acts as a scaling symmetry
on the fields, whereas $\mathfrak{n}_+ = \langle {\bf e} \rangle$ acts as a shift symmetry on $C$.
The hidden symmetries in this example are generated by $\mathfrak{n}_- = \langle {\bf f}\rangle$,
inducing a non-linear action on the scalar fields.
Let us also note, that the Lagrangian (\ref{LPP}) for this example is given by
\begin{equation}
{\cal L}_{\rm coset} =
-\frac14\,|e| \partial_\mu\phi \,\partial^\mu\phi
- \frac14\,|e|\,{\rm e}^{-2\phi} \,\partial_\mu C \,\partial^\mu C
\;,
\label{Lsl2}
\end{equation}
which is invariant under the transformations (\ref{hef}).

This is the coset space that already appears in the ten-dimensional IIB supergravity (\ref{cosetIIB})
(with $C_0=C$, $\phi\rightarrow-\phi$).
In lower-dimensional supergravities, this coset space shows up, for example,
in the reduction of $D=5$ minimal supergravity on a circle $S^1$. 
The bosonic field content of the $D=5$ theory comprises the metric and
a single vector field. 
Its $S^1$ reduction follows the scheme described in sections~\ref{subsec:reductiongravity}, \ref{subsec:reductionsupergravity},
with $\gamma=-\frac12, \beta=\frac32$. In $D=4$ dimensions, it
gives rise to gravity coupled to two vectors and two scalar fields.
According to the symmetry enhancement (\ref{gss}), the two scalars build 
an ${\rm SL}(2)/{\rm SO(2)}$ sigma model, given by (\ref{Lsl2}).
The geometric symmetries (\ref{gnss}) in this example contain the 
\begin{equation}
\mathfrak{gl}(1)=\mathfrak{g}_0=\langle{\bf h}\rangle\;,
\label{gl1sl2}
\end{equation} 
whose action
(\ref{GL1}), (\ref{actionGL1}) on the scalar fields is reproduced by $-\frac12\delta_{\bf h}$ in (\ref{hef}).
The same formulas show that the two vector fields arising in this reduction carry charges $(+\frac32)$ and $(+\frac12)$,
respectively, under $\mathfrak{gl}(1)$. Together with their dual vectors (defined via  the duality equation
$\tilde{F}_{\mu\nu}=\frac12\,|e|\,\varepsilon_{\mu\nu\rho\sigma}F^{\rho\sigma}$ in $D=4$ dimensions),
they fill the spin-$\frac32$ representation of ${\rm SL}(2)$.
This illustrates the fact the realization of the full symmetry group in general involves the original
and the dual fields of the theory.

In a similar way, the coset space ${\rm SL}(2)/{\rm SO(2)}$ appears as one of the factors of the
scalar target space in $D=8$ maximal supergravity, c.f.\ Table~\ref{table:groups}, 
as we shall discuss in more detail in section~\ref{subsec:D8765} below.

Yet another example featuring the same global symmetry group ${\rm SL}(2)$ comes from the 
$S^1$ reduction of four-dimensional Einstein gravity. In this case, the geometric scaling symmetry 
is still (\ref{gl1sl2}), but the shift symmetry $\delta_{\bf e}$ 
acts on the scalar that is obtained by dualizing the three-dimensional Kaluza-Klein vector $A_\mu$. 
The non-linear action of the hidden symmetry $\delta_{\bf f}$ in this example was originally 
discovered in~\cite{Ehlers:1957}, and this ${\rm SL}(2)$ group goes under the name of the {\em Ehlers group}.
It is instructive to spell out some details of this first example of hidden symmetries. The $S^1$ reduction of $D=4$ gravity is described by 
an ansatz (\ref{vielbeinreduced}) with $\gamma=-1$, $\beta=2$. According to (\ref{reductionG}), it results in a
three-dimensional Lagrangian\footnote
{W.r.t.\ the conventions of (\ref{vielbeinreduced}), (\ref{reductionG}), we have rescaled $\phi\rightarrow\phi/2$.}
\begin{eqnarray}
{\cal L}^{(3)} &=&
   |e| \, R_{(3)}-\frac12\,|e|\,\partial_\mu \phi \,\partial^\mu \phi
-\frac1{4}\,|e|\,{\rm e}^{2\,\phi}\,F_{\mu\nu}\,F^{\mu\nu}
\;,
\label{reduction43}
\end{eqnarray} 
which exhibits the ${\rm GL}(1)$ symmetry (\ref{GL1}), scaling the scalar and the vector field.
In three dimensions, vector fields are dual to scalar fields. This is most conveniently implemented
in the Lagrangian (\ref{reduction43}) by treating $F_{\mu\nu}$ as a fundamental field (instead of the gauge potential $A_\mu$)
and implementing its Bianchi identity by means of a Lagrange multiplier $C$ as
\begin{eqnarray}
{\cal L}^{(3)}_{\rm parent}&=&
   |e| \, R_{(3)}-\frac12\,|e|\,\partial_\mu \phi \,\partial^\mu \phi
-\frac1{4}\,|e|\,{\rm e}^{2\,\phi}\,F_{\mu\nu}\,F^{\mu\nu} -\frac12\,\varepsilon^{\mu\nu\rho}\,\partial_\mu F_{\nu\rho}\,C
\;.
\label{reduction43parent}
\end{eqnarray} 
The resulting field equations for $F_{\mu\nu}$ are algebraic and can be used to eliminate this field, leading to the dual Lagrangian
\begin{eqnarray}
{\cal L}^{(3)}_{\rm dual}&=&
   |e| \, R_{(3)}-\frac12\,|e|\,\partial_\mu \phi \,\partial^\mu \phi
-\frac1{2}\,|e|\,{\rm e}^{-2\,\phi}\,\partial_\mu C \partial^\mu C
\;,
\label{reduction43dual}
\end{eqnarray} 
in terms of two scalar fields. This is precisely the coset space sigma model (\ref{Lsl2}). After dualizing the three-dimensional vector $A_\mu$ into a scalar field, the symmetry of the Lagrangian is thus enhanced to ${\rm SL}(2)$. Uplifting its action (\ref{hef}) back to $D=4$ dimensions then induces a `hidden' symmetry of general relativity acting on solutions with a ${\rm U}(1)$ isometry.


\subsection{$D=8, 7, 6, 5$ maximal supergravities}
\label{subsec:D8765}

We have seen in section~\ref{subsec:maximalD9} that the reduction of 11D supergravity on a two-torus $T^2$
leads to maximal $D=9$ supergravity whose global symmetries (\ref{GL2GL1}) do not include any hidden
symmetries but are limited to the geometric symmetries $\mathfrak{g}_0$ in (\ref{gnss}).
Hidden symmetries appear upon reduction on larger tori.

Let us start with $D=8$ maximal supergravity, obtained by reduction of 11D supergravity on $T^3$~\cite{Salam:1984ft}. 
The necessity of a symmetry enhancement in this theory can already be deduced from simple group 
theory considerations. As discussed above, toroidal reduction of 11D supergravity on a three-torus $T^3$
yields a theory with manifest geometric symmetries (\ref{gnss})
\begin{equation}
\mathfrak{gl}(3) \oplus 1_+
\;,
\label{nss8A}
\end{equation}
where $1_+$ denotes the one-dimensional algebra $\mathfrak{n}_+$ generated by shifts (\ref{shift1}) on the scalar
descending from the 11D 3-form. On the other hand, it follows from the discussion of section~\ref{subsec:maximalD9} 
that the same $D=8$ supergravity is obtained by reducing IIB supergravity on a two-torus $T^2$.
In this case, the higher-dimensional origin implies a global symmetry group (\ref{gnss})
\begin{equation}
\Big(\mathfrak{gl}(2) \oplus \mathfrak{sl}(2) \Big)  \oplus 2_+\;,
\label{nss8B}
\end{equation}
where $\mathfrak{sl}(2)$ is the symmetry of the IIB theory and $2_+$ denotes the 
two-dimensional algebra $\mathfrak{n}_+$ generated by shifts (\ref{shift1}) on the scalars
descending from the doublet of IIB 2-forms.
Thus, the full global symmetry algebra of $D=8$ supergravity must unite both, (\ref{nss8A}) and (\ref{nss8B}).
This is realized by the group
\begin{equation}
{\rm E}_{3(3)} \simeq {\rm SL}(3) \times {\rm SL}(2) 
\;,
\label{E3}
\end{equation}
whose algebra admits the decompositions
\begin{eqnarray}
\mathfrak{e}_{3(3)} = \mathfrak{sl}(3) \oplus \mathfrak{sl}(2) 
&\stackrel{11{\rm D}}{\longrightarrow}& 1_- \oplus \mathfrak{gl}(3) \oplus 1_+
\;,\\
\mathfrak{e}_{3(3)} = \mathfrak{sl}(3) \oplus \mathfrak{sl}(2)  
&\stackrel{\rm IIB}{\longrightarrow}&  2_- \oplus  \Big(\mathfrak{gl}(2) \oplus \mathfrak{sl}(2) \Big)  \oplus 2_+
\;,
\nonumber
\end{eqnarray}
embedding both, (\ref{nss8A}) and (\ref{nss8B}), in accordance with (\ref{gss}).
With the labelling (\ref{hef_gen}) of $\mathfrak{sl}(2)$ generators, the geometric symmetries (\ref{nss8A}) from 11D supergravity
are identified as
\begin{equation}
\langle {\bf h}\rangle = \mathfrak{gl}(1)\subset\mathfrak{gl}(3)
\;,\quad
\langle {\bf e}\rangle = 1_+
\;.
\label{he_id}
\end{equation}

The scalars of $D=8$ maximal supergravity descend from the internal block of the 11D metric together with the single scalar $C$
descending from the 11D 3-form. The symmetry enhancement (\ref{E3}) corresponds to the enhancement of the coset space
(\ref{cosetGLd}) to
\begin{equation} 
{\rm GL}(3) / {\rm SO}(3)  \xhookrightarrow {}  \left({\rm SL}(3)/{\rm SO}(3)\right) \times \left({\rm SL}(2)/{\rm SO}(2) \right)
\;.
\end{equation}
In particular, the determinant of the internal metric combines with the scalar $C$ into an ${\rm SL}(2)/{\rm SO}(2)$
coset space sigma model, similar to the example of minimal $D=5$ supergravity discussed after (\ref{Lsl2}) above.
As another non-trivial consequence of the symmetry enhancement, the vector fields of $D=8$ supergravity
\begin{equation}
\{ A_\mu{}^m, A_{\mu\,mn} \}\;,\qquad m,n =1, 2, 3\;,
\end{equation}
descending from the 11D metric and 3-form, respectively, combine into an ${\rm SL}(2)$ doublet, i.e., span 
the $(3,2)$ representation of ${\rm SL}(3) \times {\rm SL}(2)$.
Indeed, one may check with (\ref{GL1}), (\ref{actionGL1}), and the identification (\ref{he_id}) 
that these fields have opposite charges $\pm\frac12$ under ${\bf h}\in\mathfrak{sl}(2)$. 
Moreover, under ${\bf e}$, acting as shift symmetry (\ref{shift1}) on the scalar $C$, 
the vectors $A_{\mu\,mn}$ transform into $A_\mu{}^m$, 
as follows from the higher-dimensional embedding (\ref{comp3form}).
The $D=8$, 2-forms $A_{\mu\nu\,m}$ have zero charge under ${\bf h}\in\mathfrak{sl}(2)$ and remain ${\rm SL}(2)$ singlets.
For the $D=8$, 3-form, the symmetry enhancement can only be made visible upon including the dual fields. 
To this end, consider the 3-form $A_{\mu\nu\rho}$ descending from the 
11D three form, together with its dual \begin{equation}
\tilde{A}_{\mu\nu\rho}= \frac16\,A_{\mu\nu\rho\,kmn}\,\varepsilon^{kmn}\;,
\end{equation}
descending from the 11D 6-form.\footnote{
Equivalently, in $D=8$ dimensions the two 3-forms are related by a first-order duality equation of the form
$\tilde{F}_{\mu_1 \dots \mu_4} = \frac1{24}\,|e|\,\varepsilon_{\mu_1 \dots \mu_8} F^{\mu_5\dots \mu_8}$, obtained by dimensional reduction of (\ref{F47}).
}
Equation
(\ref{actionGL1}) shows that they have opposite charges $\pm\frac12$ under ${\bf h}$ while (\ref{c6c3}) shows how they are 
mapped into each other under the action of~${\bf e}$.
This shows that the 3-form $A_{\mu\nu\rho}$ together with its dual, forms a doublet under the ${\rm SL}(2)$.
Once more, this illustrates that the realization of the full enhanced symmetry group in general involves the original
and the dual fields of the theory.

For the lower-dimensional maximal supergravities, the symmetry enhancement proceeds in an analogous way. 
For $D=7$ maximal supergravity \cite{Sezgin:1982gi}, the geometric symmetries enhance to an ${\rm E}_{4(4)}={\rm SL}(5)$ global
symmetry group, with the decompositions (\ref{gss}) corresponding to\footnote{
For the ${\rm SL}(d)$ groups, we use the notation $R'$ to denote the dual representation to $R$.}
\begin{eqnarray}
\mathfrak{e}_{4(4)} =\mathfrak{sl}(5) &\stackrel{11{\rm D}}{\longrightarrow}& 4'_- \oplus \mathfrak{gl}(4) \oplus 4_+
\;,\\
\mathfrak{e}_{4(4)} =\mathfrak{sl}(5) &\stackrel{\rm IIB}{\longrightarrow}&  (3',2)_- \oplus  \Big( \mathfrak{gl}(3) \oplus \mathfrak{sl}(2) \Big) \oplus (3,2)_+
\;,
\nonumber
\end{eqnarray}
respectively, depending on the higher-dimensional origin.

For $D=6$ maximal supergravity \cite{Tanii:1984zk}, the geometric symmetries enhance to an ${\rm E}_{5(5)}={\rm SO}(5,5)$ global
symmetry group, with the decompositions (\ref{gss}) corresponding to
\begin{eqnarray}
\mathfrak{e}_{5(5)}  =\mathfrak{so}(5,5) &\stackrel{11{\rm D}}{\longrightarrow}& 10'_- \oplus \mathfrak{gl}(5) \oplus 10_+
\;,\\
\mathfrak{e}_{5(5)} =\mathfrak{so}(5,5) &\stackrel{\rm IIB}{\longrightarrow}& (1,1)_{-2} \oplus (6,2)_{-1} \oplus  \Big( \mathfrak{gl}(4) \oplus \mathfrak{sl}(2) \Big) \oplus (6,2)_{+1} \oplus (1,1)_{+2}
\;,
\nonumber
\end{eqnarray}
respectively, depending on the higher-dimensional origin. The shift symmetries 
$(6,2)_{+1}$ and $(1,1)_{+2}$ in the IIB decomposition are realized on the scalars descending from the IIB 2-forms and 4-form,
respectively.

For $D=5$ maximal supergravity \cite{Cremmer:1980gs}, the geometric symmetries enhance to an ${\rm E}_{6(6)}$ global
symmetry group, with the decompositions (\ref{gss}) corresponding to
\begin{eqnarray}
\mathfrak{e}_{6(6)} &\stackrel{11{\rm D}}{\longrightarrow}& 1_{-2} \oplus 20_{-1} \oplus \mathfrak{gl}(6) \oplus 20_{+1} \oplus 1_{+2}
\;,\\
\mathfrak{e}_{6(6)} &\stackrel{\rm IIB}{\longrightarrow}& (5',1)_{-2} \oplus (10,2)_{-1} \oplus  \Big( \mathfrak{gl}(5) \oplus \mathfrak{sl}(2) \Big) \oplus (10',2)_{+1}\oplus (5,1)_{+2}
\;,
\nonumber
\end{eqnarray}
respectively, depending on the higher-dimensional origin.
The shift symmetry $1_{+2}$ is realized on the scalar descending from the 11D 6-form. Recall that this form is not present 
in the original 11D Lagrangian (\ref{LD11}). I.e., after reduction of (\ref{LD11}) to $D=5$ dimensions, 
the full scalar coset space sigma model ${\rm E}_{6(6)}/{\rm USp}(8)$ 
can only be realized after dualizing the 3-form $A_{\mu\nu\rho}$, descending from the 11D 3-form, into a scalar field.\footnote{The corresponding $D=5$ duality equation follows from dimensional reduction 
 of (\ref{F47}).}

We refer to~\cite{Cremmer:1997ct} for a systematic discussion 
of the maximal supergravities in various dimensions, 
together with their symmetries, and their eleven-dimensional origin.

\subsection{$D=4$ maximal supergravity}
\label{subsec:D=4}

Let us discuss in a little more detail the case of $D=4$ maximal supergravity. Historically, this was the first example 
of exceptional symmetry groups appearing in supergravity theories and playing an essential role in their explicit construction.
The field content of this theory is the massless ${\cal N}=8$ supergravity multiplet 
\begin{equation}
\Big\{ g_{\mu\nu}, \psi_\mu{}^i, A_\mu{}^\Lambda, \chi^{ijk}, \phi^{ijkl} \Big\}\;,\quad
i=1, \dots, 8\;,\quad \Lambda=1, \dots, 28
\;,
\label{N8multiplet}
\end{equation}
which comprises the graviton, 8 gravitinos, 28 vector fields, 56 spin-1/2 fermions, and 70 scalar fields.
The complete theory was obtained in~\cite{Cremmer:1979up} 
by dimensional reduction of the 11D supergravity on a seven-torus $T^7$ and realizing the exceptional symmetry group E$_{7(7)}$.
As discussed in section~\ref{subsec:reductiongravity}, the reduction of pure gravity 
from eleven dimensions down to $D=4$ dimensions yields a gravitational theory with seven 
abelian vector fields $A_{\mu}{}^{n}, \ n=1, \dots, 7$, and $1+27$ scalar fields, parametrizing the 
coset space ${\rm GL}(7)/{\rm SO}(7)$.
The dimensional reduction of the antisymmetric $3$-form to $D=4$ dimensions
as described in section~\ref{subsec:reductionsupergravity} gives rise to one 3-form field, 
seven 2-form fields, $\binom{7}{2}=21$ vectors and additional $\binom{7}{3}=35$ scalar fields. 
A priori, the field content thus looks quite different from 
the ${\cal N}=8$ multiplet (\ref{N8multiplet}). Including the (normalized) ${\rm GL}(1)$ charges
from (\ref{GL1}), (\ref{actionGL1}), we find the four-dimensional bosonic field content
\begin{eqnarray}
g_{\mu\nu}&:& 1_0\;,\nonumber\\
\phi &:& 1_0+27_0+35_{-2}\nonumber\\
A_{\mu} &:& 7'_{+3} + 21_{+1} \;,\nonumber\\
A_{\mu\nu} &:& 7_{+4}  \;,\nonumber\\
A_{\mu\nu\rho} &:& 1_{+7} \;,
\label{fields11D4D}
\end{eqnarray}
with the fields (other than the ${\rm GL}(7)/{\rm SO(7)}$ scalars) falling into 
linear ${\rm GL}(7)$ representations.
Reduction of the 11D Lagrangian (\ref{LD11}) yields the sum of (\ref{reductionG}) and (\ref{reductionC}),
together with the reduction of the 11D CS term. The symmetries of the resulting action  span the algebra
\begin{eqnarray}
 \mathfrak{gl}(7) \oplus 35'_{+2} 
\;,
\label{gnssD4}
\end{eqnarray}
where the $35'_{+2}$ generators induce the shift symmetries (\ref{shift1}) on the scalars descending from the 11D 3-form.
In order to make contact with the ${\cal N}=8$ supergravity multiplet, and realize the symmetry enhancement (\ref{gss}),
we need to first dualize the 2-forms $A_{\mu\nu\,m}$ into scalar fields (put equivalently, we trade them
for the corresponding scalars $A_{m_1\dots m_6}$ descending from the dual 11D 6-form).
With the associated shift symmetries (\ref{shift1}), the algebra of manifest global symmetries enhances to  (\ref{gnss})
\begin{eqnarray}
 \mathfrak{gl}(7) \oplus 35'_{+2} \oplus  7_{+4}
\;.
\label{gnssD42}
\end{eqnarray}
The non-trivial algebra structure of the charged generators directly descends from the 11D gauge algebra (\ref{gaugealgebra11D2}).

The dynamics of the resulting set of 70 scalar fields
is described by a sigma model on the coset space
\begin{equation}
{\rm G}/{\rm K}~=~{\rm E}_{7(7)}/{\rm SU}(8)~\supset~
{\rm GL}(7)/{\rm SO(7)}
\;,
\label{GKE7}
\end{equation}
according to (\ref{LPP}).
The symmetry enhancement from (\ref{gnssD42}) to $\mathfrak{e}_{7(7)}$ 
is realized along the lines discussed in the previous sections: decomposing ${\rm E}_{7(7)}$  according
to its $\mathfrak{gl}(1)$ grading, the algebra splits into the form (\ref{gss}) 
\begin{eqnarray}
\mathfrak{e}_{7(7)} &\stackrel{11{\rm D}}{\longrightarrow}& 7'_{-4}  \oplus 35_{-2} \oplus \mathfrak{gl}(7) \oplus 35'_{+2} \oplus  7_{+4}
\;,
\label{E7split11D}
\end{eqnarray}
with the generators of non-negative charge spanning (\ref{gnssD42}).
The negative grading generators correspond to the hidden symmetries that are present after reduction to four dimensions but have no 
manifest origin in 11D supergravity.
An explicit parametrization of the coset space (\ref{GKE7}) in terms of the 11D fields is given
in the triangular gauge (\ref{triangular}) as
\begin{eqnarray}
{\cal V} &\equiv& 
{\rm exp}\left[\varepsilon^{klmnpqr}  A_{klmnpq}\, t_{(+4)\,r}\right]
{\rm exp}\left[A_{kmn}\,t_{(+2)}^{kmn}\right]
V_{{\rm GL}(7)} 
\;.
\label{V56}
\end{eqnarray}
Here, $V_{{\rm GL}(7)} \in {\rm GL}(7)$ is the internal block of the 11D metric
(up to some power of its determinant), whereas the $t_{(+n)}$ refer to the E$_{7(7)}$ generators of positive grading in 
(\ref{E7split11D}). For convenience, all generators are evaluated in the fundamental $56$ representation.
Under the global symmetry $\mathfrak{e}_{7(7)}$, the vielbein transforms as  (\ref{actionGK}),
inducing a non-linear action on the scalar fields. The bosonic sector
of the theory can be formulated in terms of the symmetric, positive definite matrix
\begin{equation}
{\cal M}_{MN}={\cal V}_{M}{}^{\underline{A}} {\cal V}_{N}{}^{\underline{A}} 
\;,
\label{MVV}
\end{equation}
on which $\mathfrak{e}_{7(7)}$ acts by conjugation. In particular, the coset space sigma model (\ref{LPP}) can be written as
\begin{equation}
{\cal L}=
\frac{1}{8}\,|e|\,\partial_{\mu}({\cal M}^{-1})^{MN}\,\partial^{\mu}{\cal M}_{MN}
\;.
\end{equation}

In order to realize the global ${\rm E}_{7(7)}$ symmetry on the vector fields, we have to combine the
28 vector fields of (\ref{fields11D4D}) with their magnetic duals into the irreducible $56$ of ${\rm E}_{7(7)}$
\begin{eqnarray}
{56} &\rightarrow&
{7}_{-3}+{21}'_{-1}+{21}_{+1}+{7}'_{+3}\;.
\label{E7vectors}
\end{eqnarray}
Once more, this illustrates that the realization of the full enhanced symmetry group involves the original
and the dual fields of the theory.
The dynamics of the vector fields is described by the ${\rm E}_{7(7)}$ covariant twisted self-duality equation 
\begin{eqnarray}
{F}_{\mu\nu}{}^M &=&  -
\frac12\,|e|\,\varepsilon_{\mu\nu\rho\sigma}\,\Omega^{MN} {\cal M}_{NK}\,{F}^{\rho\sigma}{}^K
\;,
\label{dualityFN8}
\end{eqnarray}
for the abelian field strengths ${F}_{\mu\nu}{}^M =  2\,\partial_{[\mu} A_{\nu]}{}^M$, and $ {\cal M}_{NK}$ from (\ref{MVV}).
Here, $\Omega^{MN}$ is the constant 
antisymmetric E$_{7(7)}$-invariant symplectic tensor which establishes the embedding E$_{7(7)}\subset{\rm Sp}(28,28)$.
In particular, the relation
\begin{equation}
\Omega^{KL} {\cal M}_{LM}\Omega^{MN} {\cal M}_{NP} = -\delta^K{}_P
\;,
\end{equation}
is necessary for consistency of (\ref{dualityFN8}).
The full bosonic sector of maximal $D=4$ supergravity
can be compactly described by a pseudo-Lagrangian
\begin{equation}
 {\cal L}_{D=4} =|e|\, \Big( R
 +\frac{1}{48}\,\partial_{\mu}({\cal M}^{-1})^{MN}\,\partial^{\mu}{\cal M}_{MN}
-\frac{1}{8}\,{\cal M}_{MN}\,{F}^{\mu\nu \,M}{F}_{\mu\nu}{}^N \Big)
 \;,
\label{pseudoactionN8}
\end{equation}
combined with the twisted self-duality equation (\ref{dualityFN8}). 
Both, the Lagrangian (\ref{pseudoactionN8}) and equation (\ref{dualityFN8})
are manifestly  E$_{7(7)}$ covariant. However, (\ref{pseudoactionN8}) yields only
a pseudo-Lagrangian of the theory in that the twisted self-duality equation (\ref{dualityFN8})
does not follow from the variational principle but has to be imposed separately. Only its
derivative coincides with the second-order equation for the vector fields 
that is obtained by variation of (\ref{pseudoactionN8}).

A standard action principle of the theory can only be spelled out after sacrificing the manifest E$_{7(7)}$ invariance
and splitting the 56 vector fields into 28+28 as
\begin{equation}
A_\mu{}^M=\{A_\mu{}^\Lambda, \;A_{\mu\,\Lambda} \}\;.
\label{AA}
\end{equation} 
An action can then be constructed in terms of half of the fields $A_\mu{}^\Lambda$ 
considered as independent propagating (electric) fields, 
while the $A_{\mu\,\Lambda}$ are defined via (\ref{dualityFN8}) as their on-shell (magnetic) duals~\cite{Gaillard:1981rj}.
This is achieved by replacing the Maxwell term of (\ref{pseudoactionN8}) by the Lagrangian
\begin{eqnarray}
{\cal L}_{\rm vector} &=&
\frac1{4}\,|e|\,
{\cal I}_{\Lambda\Sigma}\,
F_{\mu\nu}{}^{\Lambda}
F^{\mu\nu\,\Sigma}
+
\frac1{8}\,
\varepsilon^{\mu\nu\rho\sigma}\,
{\cal R}_{\Lambda\Sigma}\,
F_{\mu\nu}{}^{\Lambda}
F_{\rho\sigma}{}^{\Sigma}
\;,
\label{Leven}
\end{eqnarray}
in terms of the 28 abelian field strengths $F_{\mu\nu}{}^\Lambda$,
with the symmetric kinetic matrices ${\cal I}_{\Lambda\Sigma}$ and
${\cal R}_{\Lambda\Sigma}$ related to the matrix ${\cal M}$ from (\ref{MVV}) as
\begin{eqnarray}
{\cal M} &\equiv&
-\left(
\begin{array}{cc}
{\cal I}+{\cal R}{\cal I}^{-1}{\cal R}&\;\;\;- {\cal R}{\cal I}^{-1}\\
- {\cal I}^{-1}{\cal R} & {\cal I}^{-1}
\end{array}
\right)
\;,
\label{defMeven}
\end{eqnarray}
in the split (\ref{AA}). In particular, ${\cal I}_{\Lambda\Sigma}$
is negative definite, such that the kinetic term in (\ref{Leven})
comes with the correct sign.
This yields a true action principle for the bosonic sector of $D=4$ maximal supergravity,
whose off-shell symmetry however is reduced to a subgroup of E$_{7(7)}$ whereas the
full E$_{7(7)}$ global symmetry is realized only on-shell. A similar pattern is observed 
in all even-dimensional maximal supergravities in $D<10$.

Different choices for the electric/magnetic split (\ref{AA}) correspond to different electric frames and 
can be related by symplectic rotation. These give rise to 
different off-shell formulations of the theory.
In particular, the off-shell symmetry group depends on the 
particular choice of the electric frame.
Choosing the 28 electric vectors to be
\begin{eqnarray}
{7}'_{+3}+{21}'_{-1}&:& \quad A_\mu^\Lambda ~=~ \{ A_\mu{}^m, A_\mu{}^{mn} \}
\;,
\label{vectors28}
\end{eqnarray}
among the 56 vectors (\ref{E7vectors}), the symmetry of the Lagrangian (\ref{Leven})
is given by the electric subgroup ${\rm SL}(8)\subset {\rm E}_{7(7)}$,
with the vectors (\ref{vectors28}) spanning its irreducible 28-dimensional representation.

Finally, the 3-form in (\ref{fields11D4D}) is non-propagating and can be consistently set to zero
in the dimensional reduction.
Strictly speaking, however, its field equations only imply that its field strength is constant and may be set
to an arbitrary value. Keeping this integration constant produces a one-parameter deformation of
the maximally supersymmetric theory~\cite{Aurilia:1980xj}. It breaks the global E$_{7(7)}$ symmetry
and closer inspection shows that this integration constant is only one component of an irreducible 912-dimensional 
representation E$_{7(7)}$~\cite{deWit:2007mt}.
Switching on other parameters within this representation leads to different
maximally supersymmetric theories which generically have non-abelian gauge groups
and matter charged under the gauge group; these deformations are the so-called 
gauged supergravities and may correspond to more complicated 
compactifications in the presence of background fluxes and/or geometric fluxes, 
see~\cite{Samtleben:2008pe} for a review. In particular, these theories include the compactification
of eleven-dimensional supergravity on the seven-sphere $S^7$, which gives rise
to a four-dimensional theory with compact non-abelian 
gauge group ${\rm SO}(8)$~\cite{deWit:1982bul}.

Let us finally note that maximal $D=4$ supergravity can of course also be obtained by reduction of the IIB theory
on a six-torus $T^6$. In this case, the decomposition (\ref{gss}) of the algebra is given by
\begin{eqnarray}
\mathfrak{e}_{7(7)} &\stackrel{\rm IIB}{\longrightarrow}& (1,2)_{-3}\oplus (15',1)_{-2} \oplus (15,2)_{-1} \oplus  
\Big( \mathfrak{gl}(6) \oplus \mathfrak{sl}(2) \Big) \nonumber\\
&&{}\oplus (15',2)_{+1}\oplus (15,1)_{+2} \oplus (1,2)_{+3}
\;,
\label{E7splitIIB}
\end{eqnarray}
and the shift symmetries $(15',2)_{+1}$, $(15,1)_{+2}$, and $(1,2)_{+3}$ are realized on the scalars
descending from the IIB 2-forms, 4-form and dual 6-forms, respectively.

\subsection{Lower-dimensional supergravities}

The symmetry enhancement and appearance of exceptional symmetry groups continues
and becomes even more intricate with maximal supergravities in lower dimensions.
For $D=3$ maximal supergravity \cite{Marcus:1983hb}, the geometric symmetries enhance to 
an ${\rm E}_{8(8)}$ global symmetry group, with the decomposition (\ref{gss}) corresponding to
\begin{eqnarray}
\mathfrak{e}_{8(8)} 
&\stackrel{11{\rm D}}{\longrightarrow}& 
8_{-3}\oplus {28}'_{-2}  \oplus 56_{-2} \oplus \mathfrak{gl}(8) \oplus 56'_{+1} \oplus  {28}_{+2} \oplus  {8}'_{+3}
\;.
\label{E8split11D}
\end{eqnarray}
The shift symmetries $56'_{+1}$, and ${28}_{+2}$ are realized on the scalars
descending from the 11D 3-form and dual 6-form, respectively.
A new feature arising in the reduction to $D=3$ dimensions is the fact that the realization of the full
symmetry algebra requires the dualization of the 8 Kaluza-Klein vector fields $A_\mu{}^m$ (\ref{vielbeinreduced})
into scalar fields $\phi_m$. We have already encountered this in the example of the ${\rm SL}(2)$ Ehlers symmetry (\ref{reduction43dual})
which is embedded as a subgroup into the ${\rm E}_{8(8)}$ of (\ref{E8split11D}).
The dual scalar fields encountered in $D>3$ supergravities have all been identified among the components of the 11D dual 6-form.
In contrast, the higher-dimensional interpretation of the $\phi_m$ is more subtle, as they should be identified with components
of the 11D `dual graviton' \cite{Hull:2000zn,West:2001as}, whose proper definition beyond the linearized theory and before 
dimensional reduction remains ambiguous.
The shift symmetries ${8}'_{+3}$ in (\ref{E8split11D}) act on these dual scalars $\phi_m$, and the full bosonic sector
of the theory is given by a gravity coupled sigma model on the coset space ${\rm E}_{8(8)}/{\rm SO}(16)$.
For completeness, let us also note the decomposition (\ref{gss}) of ${\rm E}_{8(8)}$ w.r.t.\ IIB supergravity
\begin{eqnarray}
\mathfrak{e}_{8(8)} &\stackrel{\rm IIB}{\longrightarrow}&
(7,1)_{-4}\oplus (7',2)_{-3}\oplus (35',1)_{-2} \oplus (21,2)_{-1} \oplus  
\Big( \mathfrak{gl}(7) \oplus \mathfrak{sl}(2) \Big) \nonumber\\
&&{}\oplus (21',2)_{+1}\oplus (35,1)_{+2} \oplus (7,2)_{+3}\oplus (7',1)_{+4}
\;.
\label{E8splitIIB}
\end{eqnarray}

In the reduction to $D=2$, the structures become even richer. Extrapolating the exceptional series of Lie algebras with
Dynkin diagram given by Figure~\ref{fig:dynkin} defines the infinite-dimensional algebra $\mathfrak{e}_{9(9)}$ as the
(centrally extended) loop algebra $\widehat{\mathfrak{e}_{8(8)}}$. This algebra naturally acts on infinite-dimensional representations
which are built by the infinite towers of dual scalar fields that are defined on-shell by repeatedly dualizing the 128 scalar fields that appear
in the reduction of 11D supergravity. These physical scalars together with the infinite tower of dual potentials can be cast
into the coset space ${\rm E}_{9(9)}/{\rm K}({\rm E}_9)$ with a well-defined action of the $\mathfrak{e}_{9(9)}$ symmetry algebra. 
We refer to \cite{Julia:1982gx,Julia:1981wc,Breitenlohner:1986um,Nicolai:1987kz,Nicolai:1988jb,Julia:1996nu,Nicolai:1998gi} for details.

The large global symmetry groups appearing in low dimensions have been of practical use in order to generate solutions
of the higher-dimensional theories. While we have seen that part of the global symmetries (\ref{gnss}) after toroidal reduction 
descends from the action of particular higher-dimensional diffeomorphisms and gauge transformations, 
the hidden symmetries combined in the algebra $\mathfrak{n}_-$ in (\ref{gss})
do not have a direct higher-dimensional interpretation. A solution of the higher-dimensional field equations with
a sufficient number of commuting Killing vector fields induces a solution of the lower-dimensional theory on which the action
of the full symmetry group associated with (\ref{gss}) can be explicitly computed. Lifting the result back to higher dimensions then
produces a genuinely new solution to the higher-dimensional field equations.
It is in this context of solution generating methods in Einstein gravity that hidden symmetries have first been discovered in $D=3$ reductions
\cite{Ehlers:1957,Neugebauer:1969wr,Geroch:1970nt,Kinnersley:1973xxx}, as well as in the infinite-dimensional case in $D=2$ reductions 
\cite{Geroch:1972yt,Belinsky:1971nt,Harrison:1978xxx,Maison:1978es,Hoenselaers:1979mk}, see \cite{Maison:2000fj} for a review.
The larger the group of hidden symmetries, the larger is the orbit of newly generated solutions.
In supergravity, the hidden symmetries from $D=3$ reductions have been exploited as solution generating techniques for the construction of black hole
and black ring solutions in higher dimensions, see e.g., \cite{Bouchareb:2007ax,Compere:2009zh,Chow:2014cca}.

Let us finally mention that the extrapolation to yet higher $d>9$ leads to the over-extended and very-extended
Kac-Moody algebras $\mathfrak{e}_{10}$ and $\mathfrak{e}_{11}$, respectively, which each have been conjectured 
in different context to play a fundamental role in the full 11D supergravity~\cite{West:2001as,Damour:2002cu}.

\section{Exceptional field theory}
\label{sec:exft}

We have seen in the preceding sections that the global symmetry groups of lower-dimensional maximal supergravity theories
are only partially explained by the diffeomorphism and gauge symmetries of 11D supergravity. In particular, after dimensional
reduction of (\ref{LD11}), it is only after dualization of some of the lower-dimensional fields that the full global symmetry group ${\rm E}_{d(d)}$
becomes manifest.
In this final section, we briefly review the reformulation of 11D supergravity as an exceptional field theory (ExFT)~\cite{Hohm:2013pua}. As an illustration, we focus on the example of ${\rm E}_{7(7)}$ ExFT~\cite{Hohm:2013uia}. In this formulation, dimensional reduction of the higher-dimensional theory directly leads to the Lagrangian (\ref{pseudoactionN8}) and equations of motion (\ref{dualityFN8}), in which the global exceptional symmetry ${\rm E}_{7(7)}$ is manifest.

Starting from 11D supergravity, we may perform a decomposition of fields (\ref{vielbeinreduced}) and (\ref{embedA}), as appropriate for dimensional reduction on $T^7$, and rewrite the theory in terms of the various components, without however imposing (\ref{torus_truncation}), i.e., keeping the full 
eleven-dimensional coordinate dependence of all fields. This is akin to a Kaluza-Klein compactification of the 11D theory in which all the Kaluza-Klein towers of massive fields are kept. From the four-dimensional point of view, this corresponds to a theory with infinitely many fields, compactly encoded in the dependence of all fields on the internal coordinates $y^m$. As a standard structure of Kaluza-Klein theory, the resulting theory comes with an infinite-dimensional non-abelian gauge structure, corresponding to the diffeomorphisms and gauge symmetries on the internal space.
After dualization of fields,
following the steps of section~\ref{subsec:D=4},\footnote{In the reduction to $D=4$ discussed in section~\ref{subsec:D=4}, dualization always refers to abelian vectors and $p$-forms. Here, the non-abelian structure related to the dependence on internal coordinates does not pose an obstruction to the dualization but can be compensated by St\"uckelberg-type couplings to higher degree forms, as is common in gauged supergravity \cite{deWit:2008ta,Samtleben:2008pe}. In particular, this is the underlying reason for extending the bosonic field content to (\ref{ExFTfieldsE7B}).}  this leads to a formulation of the bosonic field content in terms of the E$_{7(7)}$ fields
\begin{equation}
\left\{
g_{\mu\nu}, {\cal M}_{MN}, {\cal A}_{\mu}{}^M \right\}
\,,\qquad
\mu, \nu =0, \dots, 3\;,\quad
M=1, \dots, 56\,,
\label{ExFTfieldsE7A}
\end{equation}
in alignment with the field content of $D=4$ maximal supergravity, except for all fields keeping their dependence on the internal coordinates $y^m$.
In particular, ${\cal M}_{MN}$ still is a matrix of type (\ref{MVV}), representing the coset space E$_{7(7)}/{\rm SU}(8)$. Its vielbein (\ref{MVV}) is parametrized as (\ref{V56}) in terms of the 11D fields.
On top of these fields, the formulation requires 2-forms of the type
\begin{equation}
\left\{
{\cal B}_{\mu\nu\,\alpha}, {\cal B}_{\mu\nu\,M} \right\}
\,,\qquad
\alpha=1, \dots, 133\,,
\label{ExFTfieldsE7B}
\end{equation}
also depending on all coordinates.
Here, $\alpha$ is an index in the adjoint representation of E$_{7(7)}$, while the 56 2-forms ${\cal B}_{\mu\nu\,M}$ satisfy algebraic constraints
\begin{equation}
(t_\alpha)_K{}^M \Omega^{NK}\,{\cal B}_{\mu\nu\,M} {\cal B}_{\rho\sigma\,N} = 0 = \Omega^{MN}\,{\cal B}_{\mu\nu\,M}  {\cal B}_{\rho\sigma\,N} 
\;,
\label{sectionB}
\end{equation}
where $(t_\alpha)_K{}^M$ denote the generators of the algebra $\mathfrak{e}_{7(7)}$.

The 11D field equations can be written in terms of the objects (\ref{ExFTfieldsE7A}), (\ref{ExFTfieldsE7B}). 
Remarkably, the resulting equations can equivalently be derived from first principles based on the infinite-dimensional gauge structure of the theory.
To this end, the internal coordinates are embedded into an extended spacetime
with coordinates transforming in the fundamental 56 of E$_{7(7)}$.
The original physical coordinates are recovered as the solution of an E$_{7(7)}$-covariant section constraint. 
On the extended spacetime, the original  diffeomorphisms and gauge symmetries are unified
into generalized diffeomorphisms~\cite{Hull:2007zu,Pacheco:2008ps,Hillmann:2009pp,Berman:2010is,Coimbra:2011ky,Berman:2012vc,Coimbra:2012af,Cederwall:2013naa,Bossard:2017aae,Cederwall:2017fjm},
which provide the organizing structure for the construction of the theory. 

Specifically, the action of generalized diffeomorphisms on the scalar matrix ${\cal M}_{MN}$ is of the form \cite{Coimbra:2011ky,Berman:2012vc}
\begin{equation}
\delta_\Lambda {\cal M}_{MN}=
{\cal L}_\Lambda {\cal M}_{MN} = \Lambda^K \partial_K {\cal M}_{MN} 
+24\,\partial_L\Lambda^K\,\mathbb{P}^K{}_L{}^P{}_{(M}\,{\cal M}_{N)P}
\,.
\label{gendiff}
\end{equation}
Here, $\mathbb{P}^K{}_L{}^P{}_{M}$ is the projector on the adjoint representation of E$_{7(7)}$,
which takes the explicit form\footnote{For the raising and lowering of symplectic indices, we use North-West South-East conventions,
i.e., $Z^M=\Omega^{MN}Z_N$ and $Z_M = Z^N \Omega_{NM}$.}
\begin{eqnarray}
\mathbb{P}^K{}_M{}^L{}_N&=&
(t_\alpha)_M{}^K (t^\alpha)_N{}^L \\
&=&
\frac1{24}\,\delta_M{}^K\delta_N{}^L
+\frac1{12}\,\delta_M{}^L\delta_N{}^K
+(t_\alpha)_{MN} (t^\alpha)^{KL}
-\frac1{24} \,\Omega_{MN} \Omega^{KL}
\;.\nonumber
\label{PadjE7}
\end{eqnarray}
The 56 components of the gauge parameter $\Lambda^M$ in (\ref{gendiff}) combine the gauge symmetries associated with the 
vector fields (\ref{E7vectors}). I.e., $7+21$ of the gauge symmetries descend from the 11D diffeomorphisms and 
tensor gauge transformations, respectively,
while the other half is associated with the dual vector fields.

The coordinate dependence of all fields and gauge parameters is constrained by the so-called section constraint. The latter imposes an
embedding of the physical coordinates $\partial_m \xhookrightarrow {} \partial_M$ 
imposing that every couple of fields $(\Phi_1, \Phi_2)$ satisfies
\begin{equation}
(t_\alpha)_K{}^M \Omega^{NK}\,{\partial}_{M} \Phi_1 {\partial}_{N}\Phi_2 = 0 = \Omega^{MN}\,{\partial}_{M}\Phi_1  {\partial}_{N} \Phi_2
\;,
\label{section-constraint}
\end{equation}
similar to (\ref{sectionB}). This is an E$_{7(7)}$-covariant way of stating that only 7 out of the formally 
56 derivatives $\partial_M$ appearing in (\ref{gendiff}) 
actually have a non-trivial action, as compatible with the eleven-dimensional nature of the original theory.
It is straightforward to check that the algebra of generalized diffeomorphisms (\ref{gendiff}) only closes under the assumption of (\ref{section-constraint}).

Not unexpectedly, there are in fact two inequivalent maximal solutions to (\ref{section-constraint}) 
that restrict the dependence of all fields to 7 and 6 coordinates, respectively.
They correspond to 11D and IIB supergravity, respectively,
which are thus both embedded into the same exceptional field theory.
Specifically, the solutions to the section constraint (\ref{section-constraint}) are
based on the decompositions (\ref{E7split11D}) and 
(\ref{E7splitIIB}), respectively, of  $\mathfrak{e}_{7(7)}$
and correspond to the embedding of internal coordinates  
$\partial_m \xhookrightarrow {} \partial_M$ realized according to\footnote{
The fact that these embeddings provide solutions to (\ref{section-constraint})
can immediately be inferred from the $\mathfrak{gl}(1)$ gradings.}
\begin{eqnarray}
\mbox{11D} ~:~
\mathfrak{e}_{7(7)} &\longrightarrow& \mathfrak{gl}(7)
\nonumber\\
 {56} &\longrightarrow&
\boxed{ {7}_{-3}} +{21}'_{-1}+{21}_{+1}+{7}'_{+3}
\;,
\nonumber\\[2ex]
{\rm IIB}  ~:~
\mathfrak{e}_{7(7)} &\longrightarrow& \mathfrak{gl}(6) \oplus \mathfrak{sl}(2)
\nonumber\\
{56} &\longrightarrow&
\boxed{({6},1)_{-4}}+(6',2)_{-2}+({20},1)_0 +(6,2)_{+2}+ (6',1)_{+4}
\;.
\label{coordinates}
\end{eqnarray}
Invariance under local gauge transformations (\ref{gendiff}) in ExFT is implemented by 
covariant derivatives
\begin{equation}
{\cal D}_{\mu} =
\partial_\mu -{\cal L}_{{\cal A}_\mu}
\,,
\label{covD}
\end{equation}
with the vector fields ${\cal A}_\mu{}^M$ from (\ref{ExFTfieldsE7A}) transforming as
\begin{eqnarray}
\delta_\Lambda {\cal A}_\mu{}^{M}
&=& \partial_\mu \Lambda^M - {\cal A}_\mu{}^K \partial_K \Lambda^{M}
+12\,\partial_K{\cal A}_\mu{}^L\,\mathbb{P}^K{}_L{}^M{}_{N}\,\Lambda_\mu{}^{N}
-\frac12\,\Lambda^M \partial_K {\cal A}_\mu{}^{K}
\nonumber\\
&=& {\cal D}_\mu \Lambda^M\,.
\label{gendiffA}
\end{eqnarray}
In turn, the gauge covariant field strength is given by
\begin{eqnarray}
{\cal F}_{\mu\nu}{}^M &=&
2\, \partial_{[\mu} {\cal A}_{\nu]}{}^M 
-2\,{\cal A}_{[\mu}{}^K \partial_K {\cal A}_{\nu]}{}^M 
-12\, (t_\alpha)^{MK} (t^\alpha)_{NL}
\,{\cal A}_{[\mu}{}^N\,\partial_K {\cal A}_{\nu]}{}^L 
\label{F7}
\\
&&{}
-\frac1{2} \Omega^{MK}\,{\cal A}_{[\mu}{}^N\,\partial_K {\cal A}_{\nu]\,N}
 - 12 \,  (t^\alpha)^{MN} \partial_N {\cal B}_{\mu\nu\,\alpha}
-\frac12\,\Omega^{MN} {\cal B}_{\mu\nu\,N}
\,.
\nonumber
\end{eqnarray}
While the non-abelian part of the field strength immediately follows from the algebra of generalized diffeomorphisms (\ref{gendiff}),
the St\"uckelberg-type couplings to the 2-forms of (\ref{ExFTfieldsE7B}) are required by gauge covariance since the algebra is a
Leibniz rather than a Lie algebra.

The dynamics of E$_{7(7)}$ ExFT  is described by a twisted self-duality equation for these non-abelian field strengths
which directly generalizes the corresponding equation of the $D=4$ theory (\ref{dualityFN8})
\begin{equation}
{\cal F}_{\mu\nu}{}^M =
-\frac12\,|e|\,\varepsilon_{\mu\nu\rho\sigma}\,\Omega^{MN} {\cal M}_{NK}\,{\cal F}^{\rho\sigma\,K}
\,.
\label{twistedSDExFT}
\end{equation}
Similar to the $D=4$ theory, the remaining field equations of E$_{7(7)}$ ExFT are obtained from a pseudo-action whose Lagrangian is
directly modeled after (\ref{pseudoactionN8}) as
\begin{eqnarray}
 {\cal L}_{\rm ExFT7} &=& {|e|}\, \Big( {\cal R}
 +\frac{1}{48}\,g^{\mu\nu}{\cal D}_{\mu}({\cal M}^{-1})^{MN}\,{\cal D}_{\nu}{\cal M}_{MN}
-\frac{1}{8}\,{\cal M}_{MN}{\cal F}^{\mu\nu\, M}{\cal F}_{\mu\nu}{}^N \Big)
\nonumber\\
&&{}
 +{\cal L}_{\rm top}
-{|e|}\, V(g,{\cal M})\;,
\label{ExFTE7}
\end{eqnarray}
upon introducing an internal coordinate dependence for all fields and rendering all terms invariant under the
action of generalized diffeomorphisms (\ref{gendiff}), (\ref{gendiffA}).

Here, the Einstein-Hilbert term is constructed from the modified Ricci scalar ${\cal R}$, constructed from the external metric $g_{\mu\nu}$ 
however using covariant derivatives 
\begin{equation}
{\cal D}_{\mu} g_{\nu\rho} =  \partial_\mu - {\cal A}_\mu{}^K \partial_K g_{\nu\rho}
- \partial_K {\cal A}_\mu{}^K \, g_{\nu\rho}
\;.
\end{equation}
Similarly, the scalar kinetic term in (\ref{ExFTE7}) is a gauged sigma model with covariant derivatives defined by (\ref{covD}), (\ref{gendiff}).
The Yang-Mills term is built from the field strengths (\ref{F7}), while the 
non-abelian topological term is most compactly defined as the boundary contribution
of a five-dimensional integral over
\begin{equation}
d{\cal L}_{\rm top} \propto \Omega_{MN}\,{\cal F}^M \wedge {\cal D} {\cal F}^N \,,
\label{dLtopE7}
\end{equation}
with the covariant derivative ${\cal D} {\cal F}^N$ defined as in (\ref{gendiffA}).
Finally, the potential term $V(g,{\cal M})$ in (\ref{ExFTE7}) is given by
\begin{equation}\label{fullpotential}
 \begin{split}
  V(g,{\cal M}) \ = \ &-\frac{1}{48}{\cal M}^{MN}\partial_M{\cal M}^{KL}\,\partial_N{\cal M}_{KL}
  +\frac{1}{2} {\cal M}^{MN}\partial_M{\cal M}^{KL}\partial_L{\cal M}_{NK}\\
  &-\frac{1}{4}\,|e|^{-1}\partial_M|e|\,\partial_N{\cal M}^{MN}-\frac{1}{16}  {\cal M}^{MN}
  |e|^{-2}\partial_M|e| \partial_N|e|
\\
  &  -\frac{1}{4}{\cal M}^{MN}\partial_Mg^{\mu\nu}\partial_N g_{\mu\nu}\;. 
 \end{split} 
\end{equation}
Although not manifest, this term can be shown to be invariant under generalized 
diffeomorphisms (\ref{gendiff}) up to total derivatives. It may be thought of as the analogue of a Ricci scalar
on the extended internal spacetime~\cite{Coimbra:2011ky}.

By construction, all terms in (\ref{ExFTE7}) are separately invariant under generalized diffeomorphisms (\ref{gendiff}), (\ref{gendiffA}). Moreover, the relative coefficients among these terms are uniquely fixed by further demanding invariance under properly defined external diffeomorphisms. The latter are generated by vectors $\xi^\mu$, which however also depend on the internal coordinates. Invariance under generalized internal and external diffeomorphisms thus uniquely fixes the dynamics of E$_{7(7)}$ ExFT.

After picking a solution (\ref{coordinates}) of the section constraint, the field equations
obtained from \eqref{twistedSDExFT} and \eqref{ExFTE7} precisely reproduce the field equations
of 11D and IIB supergravity, respectively. 
Moreover, also massive IIA supergravity, c.f.\ (\ref{mIIA}), 
can be reproduced upon further deformation 
of the gauge structures~\cite{Ciceri:2016dmd,Cassani:2016ncu}.
All these theories are thus united within a common framework. Moreover, after
dimensional reduction $\partial_M\rightarrow0$ this formulation directly yields the 
$D=4$ theory (\ref{dualityFN8}), (\ref{pseudoactionN8}), with the global symmetry group E$_{7(7)}$ manifest.
This makes ExFT a natural framework for the study of duality and solution generating transformations, see e.g.,
\cite{Malek:2019xrf,Musaev:2020nrt,Blair:2022gsx}.
Moreover, with the higher-dimensional fields already re-arranged such as to fit the fields of the lower-dimensional
theory, E$_{7(7)}$ ExFT (\ref{ExFTE7}) is precisely tailored for the study of reductions to $D=4$ dimensions.
It has been a particularly powerful tool for the construction of consistent trunctions \cite{Lee:2014mla,Hohm:2014qga} 
and the computation of Kaluza-Klein spectra around four-dimensional backgrounds \cite{Malek:2019eaz}.

Exceptional field theories have been constructed for all finite-dimensional duality groups E$_{d(d)}$ (i.e., for $d\le8$) 
\cite{Hohm:2013pua,Hohm:2013vpa,Hohm:2013uia,Hohm:2014fxa,Hohm:2015xna,Abzalov:2015ega,Musaev:2015ces,Berman:2015rcc}.
Just as (\ref{ExFTE7}), the respective actions are modeled after the structure of the $(11-d)$-dimensional maximal supergravities,
lifting all fields to an extended spacetime (subject to the section contraint), with the non-abelian
gauge structure induced by the infinite-dimensional algebraic structure
of generalized diffeomorphisms. For $d>8$, the exceptional field theory based on the infinite-dimensional affine
algebra $\mathfrak{e}_{9(9)}$ has been constructed in \cite{Bossard:2018utw,Bossard:2021jix}.
Extrapolating the structures all the way to the very-extended Kac-Moody algebra $\mathfrak{e}_{11}$, a master 
formulation has been given in~\cite{Bossard:2021ebg}. This also allows to make contact with the 
E$_{11}$ conjectures of \cite{West:2001as,West:2003fc,Tumanov:2016abm} and the E$_{10}$ conjecture of~\cite{Damour:2002cu}.

Let us finally note that although we have restricted here to a discussion of the bosonic sector, the ExFT construction can be extended
to the fermionic sector in a unique way such that supersymmetry can be realized~\cite{Coimbra:2012af,Godazgar:2014nqa,Musaev:2014lna,Butter:2018bkl,Bossard:2019ksx}. 
It is however interesting to note that, as stated above, in this framework, the bosonic sector is already uniquely determined by imposing purely bosonic symmetries, the generalized internal and external diffeomorphisms. This of course is directly related to the observation discussed after (\ref{gss}) that the appearance of hidden symmetries relies on the exact bosonic couplings determined by supersymmetry. Why supersymmetry precisely selects 
the couplings that give rise to the symmetry enhancement and the exceptional groups still remains somewhat of a mystery
and continues to challenge our understanding of the fundamental symmetries of supergravity.

\vspace*{-.2cm}

\begin{acknowledgement}
I wish to thank all my collaborators and in particular Bernard de Wit, Olaf Hohm, Emanuel Malek, Hermann Nicolai, 
Ergin Sezgin, and Mario Trigiante, for the numerous exciting discussions and collaboration on topics related to this review.
\end{acknowledgement}

\vspace*{-.7cm}


%

\end{document}